\documentclass[a4paper,onecolumn,11pt]{quantumarticle}
\pdfoutput=1
\usepackage[utf8]{inputenc}
\usepackage[english]{babel}
\usepackage[T1]{fontenc}
\usepackage{amsmath}
\usepackage{hyperref}

\usepackage{subcaption}

\usepackage{tikz}
\usepackage{lipsum}

\usepackage{cite}
\usepackage{amsmath,amssymb,amsfonts}
\usepackage{algorithmic}
\usepackage{graphicx}
\usepackage{textcomp}
\usepackage{xcolor}
\usepackage{multicol}
\usepackage{hyperref}
\hypersetup{colorlinks=true, % colorise les liens
breaklinks=true, % retours à la ligne pour les liens trop longs
urlcolor= black, % couleur des hyperliens
linkcolor= black,% couleur des liens internes aux documents 
citecolor= black % couleur des liens vers la biblio
} % comment or uncomment to colorbox the hyperref links !

\usepackage{float} % needed only to use H
\makeatletter% because def contain @ 
    \def\fps@figure{H}%tb
    \def\fps@table{h}
\makeatother

\usepackage{cleveref}
\crefname{figure}{fig.}{fig.}
\Crefname{figure}{Fig.}{Fig.}
\crefname{equation}{}{}

\usepackage{braket}
\usepackage{standalone}
\standaloneconfig{mode=image}

\usepackage{acro}
\DeclareAcronym{qa}{short=\textsc{QA}, long=Quantum Arithmetic}
\DeclareAcronym{qsp}{short=\textsc{QSP}, long=Quantum Signal Processing}
\DeclareAcronym{qft}{short=\textsc{QFT}, long=Quantum Fourier Transphorm}

\DeclareAcronym{qpu}{short=\textsc{QPU}, long=Quantum Processing Unit}
\DeclareAcronym{qsvt}{short=\textsc{QSVT}, long=Quantum Singular Value Transformation}
\DeclareAcronym{gas}{short=\textsc{GAS}, long=Grover's Adaptative Search}
\DeclareAcronym{ga}{short=\textsc{GA}, long=Grover's search Algorithm}
\DeclareAcronym{bop}{short=\textsc{BOP}, long=Binary Optimization Problem}
\DeclareAcronym{vqa}{short=\textsc{VQA}, long=Variationnal Quantum Algorithm}
\DeclareAcronym{qaoa}{short=\textsc{QAOA}, long=Quantum Approximate Optimization Algorithm}
\DeclareAcronym{nisq}{short=\textsc{NISQ}, long=Noisy Intermediate Scale Quantum computer}
\DeclareAcronym{ftqc}{short=\textsc{FTQC}, long=Fault Tolerant Quantum Computer}
\DeclareAcronym{ca}{short=\textsc{CA}, long=Cold Atoms}
\DeclareAcronym{qubo}{short=\textsc{QUBO}, long=Quadratic Unconstrained Binary Optimization}
\DeclareAcronym{hubo}{short=\textsc{HUBO}, long=High order Unconstrained Binary Optimization}
\DeclareAcronym{hobo}{short=\textsc{HOBO}, long=High Order Binary Optimization}
\DeclareAcronym{cpbo}{short=\textsc{CPBO}, long=Constrained Polynomial Binary Optimization}
\DeclareAcronym{qspbo}{short=\textsc{QSPBO}, long=Quantum Signal Processing Based Oracle}
\DeclareAcronym{qabo}{short=\textsc{QABO}, long=Quantum Arithmetic Based Oracle}
\DeclareAcronym{qdbo}{short=\textsc{QDBO}, long=Quantum Dictionary Based Oracle}
\DeclareAcronym{hs}{short=\textsc{HS}, long=Hamiltonian Simulation}
\DeclareAcronym{ph}{short=\textsc{PH}, long=Problem Hamiltonian}
\DeclareAcronym{aa}{short=\textsc{AA}, long=Amplitude Amplification}
\DeclareAcronym{lcu}{short=\textsc{LCU}, long=Linear Combination of Unitaries}
\DeclareAcronym{qpe}{short=\textsc{QPE}, long=Quantum Phase Estimation}
\DeclareAcronym{qse}{short=\textsc{QSE}, long=Quantum Signal Estimation}
\DeclareAcronym{qso}{short=\textsc{QSO}, long=Quantum Signal Oraclization}
\DeclareAcronym{fpgs}{short=\textsc{FPGS}, long=Fixed-Point Grover Search}
\DeclareAcronym{qpp}{short=\textsc{QPP}, long=Quantum Phase Processing}
\DeclareAcronym{gqsp}{short=\textsc{GQSP}, long=Generalized Quantum Signal Processing}
\DeclareAcronym{aqe}{short=\textsc{AQE}, long=Ancilla Quantum Encoding}
\DeclareAcronym{hhl}{short=\textsc{HHL}, long=Harrow–Hassidim–Lloyd algorithm}
\DeclareAcronym{be}{short=\textsc{BE}, long=Block-Encoding}
\DeclareAcronym{qae}{short=\textsc{QAE}, long=Quantum Amplitude Estimation}
\DeclareAcronym{eaa}{short=\textsc{EAA}, long=Exact Amplitude Amplification}
\DeclareAcronym{qml}{short=\textsc{QML}, long=Quantum Machine Learning}

\usepackage{tikz}
\usetikzlibrary{positioning,arrows,calc,math,angles,quotes}
\usepackage{multirow}

\def\orgadep{CEA, List, F-91120}%Laboratoire Intégration des Systèmes et Technologies (LIST)} % dept. name of organization (of Aff.)
\def\loc{Palaiseau, France} % City, Country
\def\univ{Université Paris-Saclay}

\begin{document}

\title{\acl{qa}-based on \acl{qsp}}

\author{Robin OLLIVE}
\affiliation{\textit{\univ}, \orgadep, \loc}
\orcid{0009-0006-7539-363X}
\author{Stephane LOUISE}
\orcid{0000-0003-4604-6453}
% \thanks{You can use the \texttt{\textbackslash{}email}, \texttt{\textbackslash{}homepage}, and \texttt{\textbackslash{}thanks} commands to add additional information for the preceding \texttt{\textbackslash{}author}. If applicable, this can also be used to indicate that a work has previously been published in conference proceedings.}
\maketitle

\begin{abstract}
    As in classical reversible computing, \acl{qa} is typically seen as a set of tools that process binary data encoded into a quantum register to set the value of another quantum register.
    This article presents another approach to explain the \acl{qa} in quantum computing.
    Here, \acl{qa} is addressed with a matrix processing point of view.
    \acl{qa} is a convenient way to construct the Query that implements the mathematical problem of interest.
    This approach is not only an interpretation with the matrix approach to encode the problem of interest; it also allows us to derive a new original technique to construct \acl{qa} with the framework of embedded \ac{qsp}.
    This work uses the link between the eigenstate amplitude and the operator phase to transform the \ac{qsp} processed amplitude into binary value extracted by \ac{qpe}.
    The explanations allowing the \ac{qsp} based \acl{qa} construction let appear natively sub-routines and functions used by well-known algorithms such as \ac{aqe}'s \ac{hhl} and \ac{qae}.
    Methods to implement this circuit are presented in the paper.
\end{abstract}

\newpage
\tableofcontents
\newpage

\section{Introduction}
Many quantum algorithms need \acl{qa} as a subroutine.
Shor's algorithm \cite{shor_algorithms_1994}, certain Grover Oracles \cite{grover_fast_1996}, and one of the \ac{hhl} \cite{harrow_quantum_2009} sub-routines rely on \acl{qa} \cite[Table.11]{wang_comprehensive_2024}.
The ability to compute functions of binary numbers is required to implement or process many classical problems in a quantum computer.
Many quantum algorithms process a superposition of binary numbers, which allows the associated problem to be solved efficiently.

\acl{qa} is often constructed with two strategies.
These strategies are based on either the binary logic 
bitwise adders \cite{orts_review_2020} assembly \cite{vedral_quantum_1996,beckman_efficient_1996} or the \ac{qft} phase adders \cite{draper_addition_2000} assembly \cite{ruiz-perez_quantum_2017, sahin_quantum_2020, yuan_improved_2023,atchade-adelomou_efficient_2023}\footnote{A recent approach proposes to construct a more efficient \ac{qft}-based multiplier directly \cite{ramezani_quantum_2023}; nevertheless, a deep analysis of the sub-routines seems to cancel this advantage \cite{pfeffer_comment_2024}.}.
Both these strategies combine these simple blocks to elaborate complex functions.
It can lead to polynomial approximations of arbitrary functions \cite{haner_optimizing_2018}.
Many developments to improve, categorize and evaluate them exist in the literature \cite{hansen_resource_2023, wang_comprehensive_2024}.
This part of the algorithm has an extensive cost of quantum resources due to the high number of required ancilla qubits.
These ancilla qubits are needed to store the intermediate results \cite{haner_optimizing_2018}.
The trade-off between the number of ancilla qubits and the number of operations is one of the key aspects in designing a quantum circuit for \acl{qa}.

Recent developments in quantum algorithms allow to interpret the implemented problem as qubitized matrices and bring a routine, the \ac{qsp} \cite{yoder_fixed-point_2014, low_methodology_2016, low_optimal_2017,low_hamiltonian_2019, martyn_grand_2021}, that competes with \ac{hhl} to process the matrices.
The \ac{qsp} and its generalization, the \ac{qsvt} \cite{gilyen_quantum_2019-1} routine allow the processing of real part of matrix amplitudes.
\ac{qsp} does polynomial functions of the amplitudes with the polynomial order equal to the number of matrix repetitions \cite{martyn_grand_2021}.
Composing layers of \ac{qsp} and therefore function composition is possible using the more restrictive embedded \ac{qsp} or \ac{qsvt} \cite{rossi_semantic_2023}.
It gives access to more complex functions.
Using these tools in order to construct \acl{qa} with a smooth function process thanks to \acl{qsp} is the core idea of this paper.
The idea of avoiding \acl{qa} to construct a function was already developed in \cite{mcardle_quantum_2022} to prepare a quantum state.

This article proposes two contributions.
The first contribution is to show the links between \acl{qa} and the other techniques of problem implementation.
It explains the \ac{qft}-based adder structure directly.
This part shows why it implies that projector constructs from \acl{qa} are a particular \ac{hhl} instance.
The second contribution is an original strategy to construct the \acl{qa} routines.
This strategy is not based on a combination of simple operations. 
Instead, thanks to well-tuned \ac{qsp} techniques, the function of the binary register is computed in the amplitudes of the ancilla qubit.
Depending on the reason why the \acl{qa} is constructed, this amplitude can be read into a third register thanks to a \ac{qpe} variant, or directly transformed into a Grover oracle, or state initialization.
We give an example with the complete set of parameters that allow the implementation.
Depending on the function of the binary register, this method can be more efficient, especially regarding the number of ancilla qubits.

\section{\acl{qa} based query}
To solve problems with quantum algorithms, the first thing that must be checked is whether or not the classical problem can be formulated with quantum information.
If so, then this quantum information must be implemented as a quantum unitary gate, the query, which can be processed by the \ac{qpu}.
Two very popular queries are \acl{hs} eq.~\Cref{c_une_hs} and \acl{be}\footnote{
    Different types of \acl{be} exist.
    As soon as the matrix of interest is probabilistically applied to the vector of interest with a control register, the gate is a \acl{be}.
    The example of this paper is a specific case called signal operator in the $Y$-convention due to his behavior when called in a \ac{qsp} algorithm.} eq.~\Cref{c_un_be}.
A third one, that encodes the same quantum information is the results from a \ac{qpe} algorithm eq.~\Cref{c_un_qsp_res}.
It is often more complicated to construct it.
The \ac{qpe} result is often used to read the information because it contains the binary representation of the matrix eigen-phases $\lambda_{i}$ with the associated eigenstate $\ket{\lambda_{i}}$, which can easily be read.

\begin{equation}
    \widehat{V}_{H} = e^{i 2 \pi \frac{1}{\lambda_{max}} \widehat{H}}
    \label{c_une_hs}
\end{equation}

\begin{equation}
    \widehat{W_{y}}_{H} =
    \begin{bmatrix}
        \widehat{H} & - \sqrt{\widehat{I} - \widehat{H}^{2}} \\
        \sqrt{\widehat{I} - \widehat{H}^{2}} & \widehat{H} 
    \end{bmatrix}
    \begin{matrix}
        \ket{\psi} \ket{\psi_{\parallel}} \\
        \ket{\psi} \ket{\psi_{\perp}}
    \end{matrix}
    \label{c_un_be}
\end{equation}

with $\ket{\psi_{\parallel}}$ and $\ket{\psi_{\perp}}$ some control states.

\begin{equation}
    \widehat{QPE}_{H} = \sum_{i} \ket{{\lambda_{i}} %/ {|\lambda_{i}|}
    } \bra{{\lambda_{i}}% / {|\lambda_{i}|}
    } \otimes \ket{\mathrm{bin}[(b + \lambda_{i}) \mathrm{mod}[2^{F}]]} \bra{\mathrm{bin}[b]}
    \label{c_un_qsp_res}
\end{equation}

\begin{figure}[tb]
    \begin{center}
        %\resizebox{\linewidth}{!}{
            \includegraphics{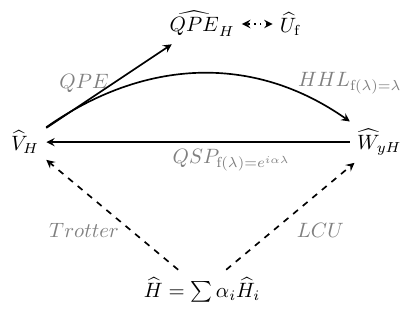}%}
    \end{center}
    \caption{Conversion between the different Query used to process information into \ac{qpu}.
    The dotted and dashed lines represent the classical data implementation, while the filled lines represent data processing on the \ac{qpu}.}
    \label{tickz_query_link}
\end{figure}

Since all these queries contain the same information, switching from query to query with another query assembly \Cref{tickz_query_link} is possible.

Another way to load the problem's data into the \ac{qpu}, the \acl{qa} \Cref{c_un_qa}, can be used for binary information:

\begin{equation}
    \begin{aligned}
        & \widehat{U}_{\mathrm{f}} = \sum_{x = 0}^{2^{F}} \ket{\mathrm{bin}[x]} \bra{\mathrm{bin}[x]} \otimes \ket{\mathrm{bin}[(b + \mathrm{f}(x))\mathrm{mod}[2^{F}]]} \bra{\mathrm{bin}[b]} \\
        & \widehat{U}_{\mathrm{f}} \ket{\mathrm{bin}[x]} \ket{0} = \ket{\mathrm{bin}[x]} \ket{\mathrm{bin}[\mathrm{f}(x)\mathrm{mod}[2^{F}]]}
    \end{aligned}
    \label{c_un_qa}
\end{equation}

While \acl{hs} and \acl{be} are based on a decomposition of the matrix formulation of the problem on a family of matrices accessible by the quantum algorithm (\Cref{tickz_query_link}, dashed lines), \acl{qa} is based on the assembly of basic logical operations.
Using the Toffolly gate and X gate, the universal gate of reversible classical logic, all the logical operations can be implemented similarly to the classical logic circuit \Cref{table_basic_logic_gate_mini} \cite[Section 3]{vedral_quantum_1996}.

\begin{table}[tb]
    \caption{Classical logic gates with a quantum computer. \\ The $S$ qubit begin in the $\ket{0}$ state.}
    \begin{center}
        % \resizebox{\linewidth}{!}{
            \includegraphics{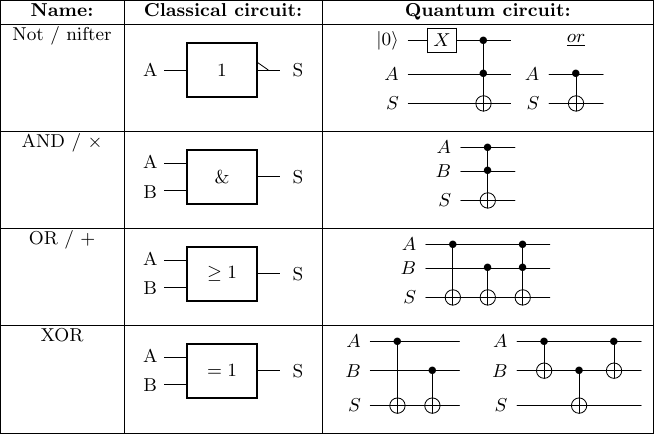}%}
    \end{center}
    \label{table_basic_logic_gate_mini}
\end{table}

The goal of this section is to show that \acl{qa} has the same shape as a \ac{qpe} result (\Cref{tickz_query_link}, dotted lines).
It will also be shown later that the two other aforementioned queries can be constructed from this \ac{qpe} result to encode a mathematical function.

\subsection{\acl{hs}-based \acl{qa}}
\label{hs_based_qa}
Let us start with an operator $\widehat{H}_{\mathrm{f}}$ whose application on a binary state associates to this state an amplitude equal to the function image of the binary value:

\begin{equation}
\begin{aligned}
    \widehat{H}_{\mathrm{f}} \ket{\mathrm{bin}[x]} & = \mathrm{f}(x) \ket{\mathrm{bin}[x]} \\
    \Rightarrow \widehat{H}_{\mathrm{f}} & = \sum_{x = 0}^{2^{F}} \mathrm{f}(x) \ket{\mathrm{bin}[x]} \bra{\mathrm{bin}[x]}
\end{aligned}
\end{equation}

In order to transform this operator into a quantum gate, the normalized\footnote{The normalization guarantees a bijective function over the interval.} \acl{hs} of this operator is implemented:

\begin{equation}
\begin{aligned}
    \widehat{V}_{\mathrm{f}} & = e^{i 2 \pi \frac{1}{\mathrm{max}[\mathrm{f}]} \widehat{H}_{\mathrm{f}}} \\
    \Rightarrow \widehat{V}_{\mathrm{f}} & = \prod_{x = 0}^{2^{F}} e^{i 2 \pi \frac{\mathrm{f}(x)}{\mathrm{max}[\mathrm{f}]}} \ket{\mathrm{bin}[x]} \bra{\mathrm{bin}[x]}
\end{aligned}
\end{equation}

This gate contains the same quantum information as the \acl{qa} associated with this function (for a function exactly computable \textit{i.e.} with enough qubits on the output register).
It is interesting to note that as for encoding an optimization problem, this gate is a diagonal matrix on the computational basis.
The \acl{qa} can be recover thanks to a \ac{qpe}:

\begin{equation}
\begin{aligned}
    \widehat{QPE}_{V_{\mathrm{f}}} \ket{\mathrm{bin}[x]} \ket{0} & = \ket{\mathrm{bin}[x]} \ket{\mathrm{bin}[\mathrm{f}(x)]} \\
    \widehat{QPE}_{V_{\mathrm{f}}} & \Leftrightarrow \widehat{U}_{\mathrm{f}} 
\end{aligned}
\end{equation}

Provided that the function \acl{hs} is accessible, it thus gives an original strategy to produce the \acl{qa} routine.
It is already used to construct \ac{qft}-based adders but currently not for more complex functions.
The generalization is developed during the second part of the paper.

\subsection{\acl{qa}-based \acl{be} and projector as specific HHL}
\label{section_hhl}
\begin{figure}[tb]
    \begin{center}
        % \resizebox{\linewidth}{!}{
            \includegraphics{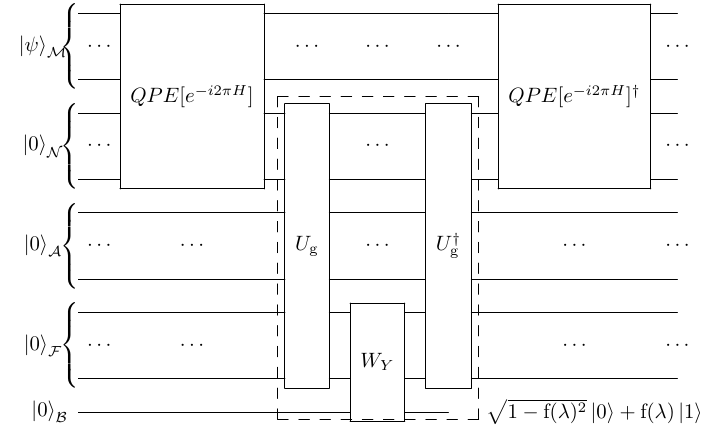}%}
    \end{center}
    \caption{\ac{hhl} quantum circuit, the dashed box: the \ac{aqe}, can be understood as making the same function on a matrix diagonal in the computational basis.
    $W_{Y}$ is used as a short notation for $C^{F}\{ \ket{\mathrm{bin}[x]} \}R_{Y}(x)$ which can be constructed as \Cref{qc_ry_kickback}.}
    \label{qc_hhl}
\end{figure}

An important utilization of the \acl{qa} is to process quantum information (eigenstate amplitudes) with a function (classical information) thanks to the \ac{hhl} algorithm \Cref{qc_hhl}.
This kind of queries implements properties of the function by selectively altering some states to apply a function on the amplitude of the states of interest.
To do so, the \acl{qa} is called in a sub-routine called \ac{aqe} as illustrated in the dashed box of \Cref{qc_hhl}.
The \acl{qa} computes the function composition of $ \mathrm{g} = \mathrm{p2a} \circ \mathrm{f} $, with $\mathrm{f}$ the target matrix function and $\mathrm{p2a}$ a specific arcsin function\cite{haner_optimizing_2018}, detailled in \Cref{section_p2a}, that correct the $\sin(x) - x$ error of the \acl{be}.
In \cite{mcardle_quantum_2022}, a similar function composition is used to associate the amplitudes function depending on binary inputs.

\begin{figure}[tb]
    \begin{center}
        % \resizebox{\linewidth}{!}{
            \includegraphics{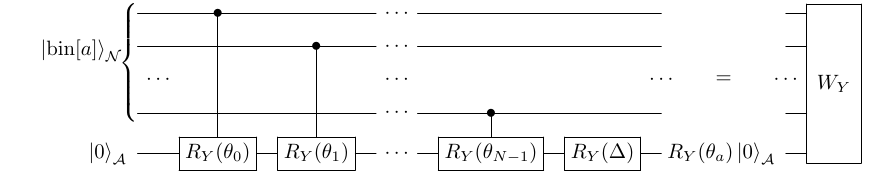}%}
    \end{center}
    \caption{$Y$-rotationnal-gate kickback also refered as the $ C^{F}\{ \ket{\mathrm{bin}[x]} \}R_{Y}(x) $ gate.
    This circuit is a $\widehat{W_{Y}}$ signal operator.
    In this diagram, $ \theta_{i} = 2 \pi \frac{2^{i}}{\alpha 2^{N}} $ and $ \Delta = 2 \pi \frac{\delta}{\alpha} $ and $ \theta_{a} = 2 \pi \frac{a}{\alpha 2^{N}} $.}
    \label{qc_ry_kickback}
\end{figure}

If the interpretation of the previous section is injected in the \ac{hhl} algorithm, it directly leads to \acl{qa}-based query.
As the \acl{qa} can be understood as the \ac{qpe} result of a diagonal matrix, the \ac{hhl} without \ac{qpe} can be understood as the matrix function of a matrix which is diagonal in the computational basis.

\begin{figure}[tb]
    \begin{center}
        % \resizebox{\linewidth}{!}{
            \includegraphics{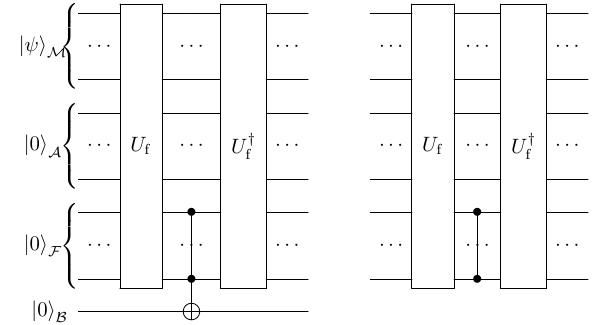}%}
    \end{center}
    \caption{Projector (1) and Grover oracle (2).
    The $n$-controlled gate identifies the state $\ket{\mathrm{bin}[y]}$ so that the qubit of the register $\mathcal{B}$ is in $\ket{1}$ for the component of $\psi$ so that $\mathrm{f}(\mathrm{bin}[\psi]) = y$.
    Instead of switching the $\mathcal{B}$ qubit, the oracle switches the sign of the corresponding state.}
    \label{qc_proj_oracle}
\end{figure}

A very popular \acl{qa}-based query is the projector \Cref{qc_proj_oracle}.
The projector corresponds to an \ac{hhl} whose associated function is a specific gate function located on the states selected thanks to \acl{qa}.
This Well-functions \ac{aqe} can be reduced to a multi-controlled gates.
This multi-controlled gate applied on the function images register allows to select the state of interest.
A Grover's oracle \Cref{qc_proj_oracle}, which marks the state identified by the projector, can easily be constructed by switching the multicontroled-NOT gates by multicontroled-phase gates.

Another query of interest: the \acl{be} of the $\widehat{H}_{\mathrm{f}}$: $\widehat{W}_{y H_{\mathrm{f}}}$ is construct if the \ac{hhl} function is the identity function: $ \mathrm{f}(x) = x $.
It is a general property, \ac{hhl}, with the identity function constructs the \acl{be} associated hermitian matrix \acl{be}.

\section{\acl{qsp}-based \acl{qa}}
This section describes how to implement \acl{qa} from the \ac{qsp} phase angles instead of many logical operations.

\subsection{QFT-based adder}
The first step to achieve \acl{qsp}-based \acl{qa}, is the \ac{qft}-based adder, which is already used to sum values due to phase additivity.
It is the opposite of the phase kickback technique in which binary logic is used to construct a phase: it uses phases to construct binary logic, \cite{cleve_quantum_1998, ossorio-castillo_generalisation_2023} .
In the literature, \ac{qft}-based adders are used as a basic logic operation to construct more complex functions.
It explains the phase additivity properties; one or many values can be added to a binary register.

\subsubsection{\acl{qpe} Interpretation of the \acl{qft} Adder}
\begin{figure}[tb]
\begin{subfigure}{0.5\textwidth}
    \begin{center}
        \resizebox{\linewidth}{!}{
            \includegraphics{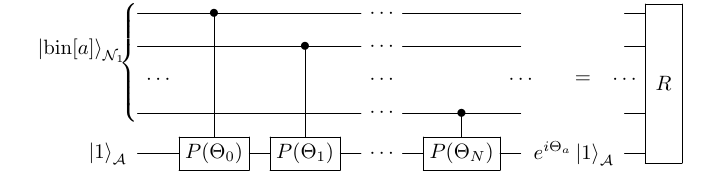}}
    \end{center}
    \vspace{0.5 cm}
    \caption{Phase Rotation matrix quantum circuit.
    The control phase gate is symmetric, so the control side can be switched.
    In this diagram, $ \Theta_{i} = \pi \frac{\alpha 2^{i}}{2^{F}} $ and $ \Theta_{a} = \pi \frac{\alpha a}{2^{F}} $.}
    \label{qc_phase_matrix}
\end{subfigure}
\begin{subfigure}{0.5\textwidth}
    \begin{center}
        \resizebox{\linewidth}{!}{
            \includegraphics{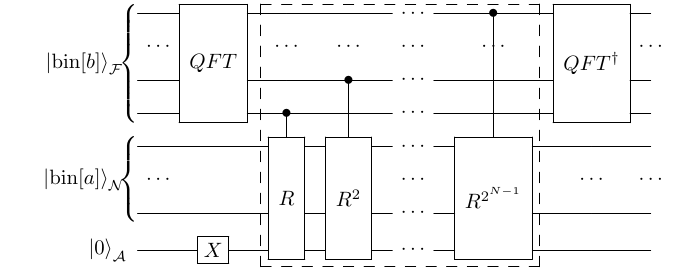}}
    \end{center}
    \caption{Phase reading quantum circuit. The dashed box contains the phase kickback of the phase matrix.
    At the end of the quantum circuit, the $\mathcal{F}$ register ends in the $\ket{\mathrm{bin}[a + b]}$ state.}
    \label{qc_qft_qpe}
\end{subfigure}
\end{figure}

\begin{figure}[tb]
\begin{subfigure}{0.5\textwidth}
    \begin{center}
        \resizebox{\linewidth}{!}{
            \includegraphics{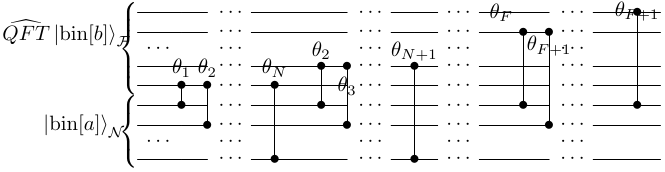}}
    \end{center}
    \caption{Reordered cleaned phase kickback quantum circuit.
    Note that (since $\theta$ is modulo $2 \pi$): $ \theta_{F + m} = \pi 2^{m} \mathrm{mod}[2 \pi] = 0 $.
    All the lines for which $\mathcal{F}$ label $>$ $\mathcal{N}$ label goes from $\theta_{N}$ to $\theta_{F}$ ($F - N$ terms).}
    \label{qc_qft_qpe_clean}
\end{subfigure}
\begin{subfigure}{0.5\textwidth}
    \begin{center}
        \resizebox{\linewidth}{!}{
        \includegraphics{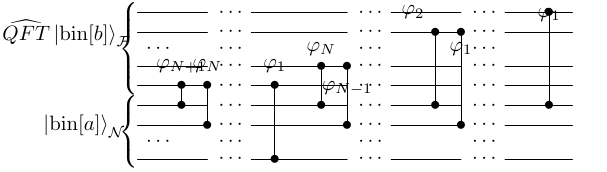}}
    \end{center}
    \caption{Other notation for the phase adder: $ \varphi_{j} = \pi \frac{1}{2^{j-1}} = \theta_{F + 1 - j} $. Here, we choose $F = N + 1$, which is enough for a simple addition. \\}
    \label{qc_qft_qpe_usu}
\end{subfigure}
\end{figure}

The \ac{qft}-based quantum adder was introduced as a well-adjusted frequency reading \cite[Section 4]{draper_addition_2000, ruiz-perez_quantum_2017} or with quantum dictionary \cite{gilliam_foundational_2021}.
This section aims to evince it as a matrix phase reading thanks to the \ac{qpe} algorithm.
The matrix of interest is diagonal with the $ 2^{N} $ phases associated with the corresponding binary values.
Thanks to the controlled-phase-adding, only the values stored in the binary register has a non-zero amplitude and are thus added, Cf. \Cref{qc_phase_matrix}.
Then, the phase reported on this qubit can be read thanks to the \ac{qpe} algorithm, \Cref{qc_qft_qpe}.
The qubit on which the phase is reporteded being an ancilla, it can be withdrawn.
Therefore, the two interpretations lead to the same quantum circuit \Cref{qc_qft_qpe_clean} \& \Cref{qc_qft_qpe_usu}.

\subsection{\acl{qsp}-based \acl{qa}}
\label{mastermind}
\begin{figure}[tb]
    \begin{center}
        \resizebox{8 cm}{!}{
            \includegraphics{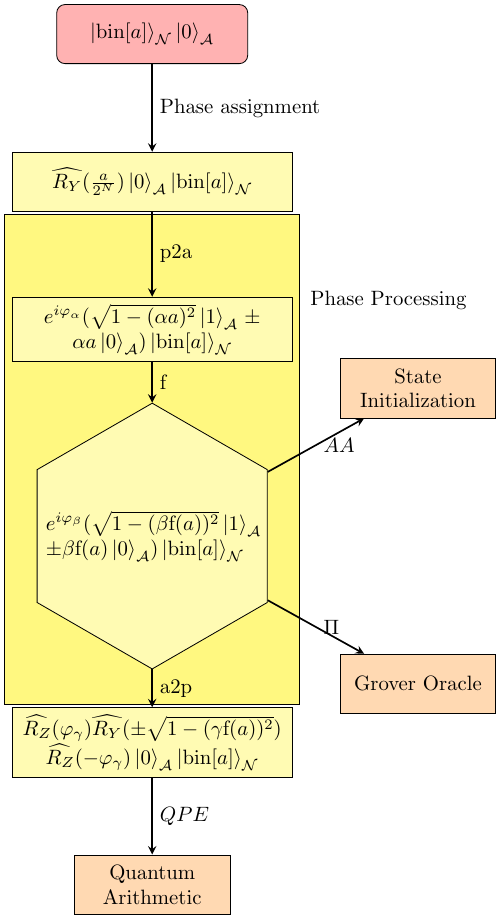}}
    \end{center}
    \caption{Workflow: \acl{qa} from embedded \ac{qsp} and \ac{qpe}.
    $\alpha$, $\beta$ and $\gamma$ are normalization factors.
    $\Pi$ referred to as a gate function and \acs{aa} refer to as \acl{aa}.}
    \label{workflow_full}
\end{figure}

The \ac{qpe} interpretation of the \ac{qft} adder tells us that other matrices carrying the phase of interest can be chosen.
Processing the phase before reading it can be advantageous to construct \acl{qa}.
The following subsection combines existing tools allowing the processing of the phase of specific matrices, \Cref{workflow_full}.
Three main steps are needed to construct the routine (called \acs{qse}), which are presented in a top-down approach:
\begin{enumerate}
    \item Assign an entangled phase proportional to the binary representation of each register.
    \item Process the phase value with the function of interest.
    \item Either read the phase, amplify or mark the result of interest to produce a useful quantum subroutine.
\end{enumerate}

The more challenging part is to apply a function to the phases.
Three strategies based on \ac{qsp} variants can be used to construct the function image:
\begin{itemize}
    \item The \ac{qsp} semantic embedding which allows to do function composition.
    \item The usual \ac{qsp} directly with the function composition: $ \mathrm{a2p} \circ \mathrm{f} \circ \mathrm{p2a} $ whose phase angles may be hard to find coupled with a \ac{qae} to measure the resulting amplitudes.
    \item Process directly the quantum phases with an usual \ac{qsp} (simplified \ac{qpp}) \cite{wang_quantum_2023}.
\end{itemize}
The \ac{qsp} semantic embedding strategy is detailed in this article.
It can be developed in four stages according to \Cref{workflow_full}:
\begin{enumerate}
    \item \acl{be} amplitudes corresponding to a rotation with a phase proportional to the register's value.
    \item Transform the \acl{be} amplitudes in values proportional to the register's value thanks to embedded \ac{qsp}.
    \item Transform this amplitude in their function images thanks to embedded \ac{qsp}.
    This routine is very useful for constructing state preparation, as proposed by \cite{mcardle_quantum_2022}.
    \item Converting these amplitudes into phases proportional to these amplitudes of the Bloch-sphere thanks to embedded \ac{qsp}.
\end{enumerate}

\subsubsection{Illustrative Example}
\begin{figure}[tb]
    \begin{center}
        \resizebox{10 cm}{!}{
            \includegraphics{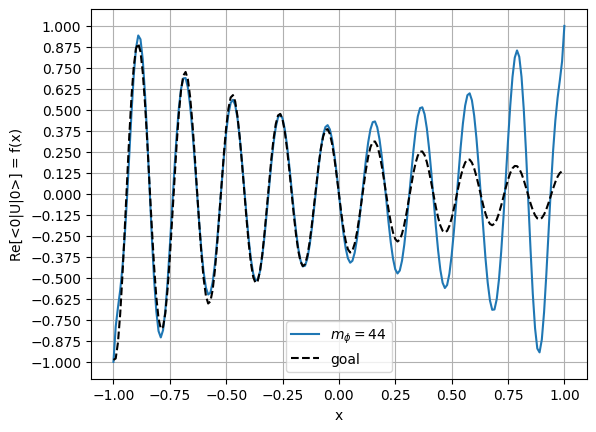}}
    \end{center}
    \caption{Full representation of the function that will be processed thanks to \ac{qse} algorithm variants.}
    \label{exemple_function_qse}
\end{figure}

The following sections illustrate the process \Cref{workflow_full} with the following function \Cref{exemple_function_qse}: $$ \mathrm{f}(x) = - \cos(15 x) e^{-x} $$
To implement it, the following function: $$ \sin(30 x) e^{- (1 + x)} $$ is prepared on a larger space.
Using a larger space allows to eliminate parity constraints of the embedded \ac{qsp} at the cost of a higher polynomial approximation order.
This example of the \acl{qse} technique encodes this function in the matrix, amplitudes, then phases, to construct a \acl{qa} routine or a Grover oracle.

\subsubsection{Binary number to Quantum Amplitudes}
\label{section_p2a}
To be able to process the quantum phases thanks to embedded \ac{qsp}, the first objective is to construct an operator that associates to each register's binary representation an amplitude whose real part is proportional to the register's value.

This ability is also needed in the HHL algorithm and is often called \ac{aqe}.
The simplest way to achieve it is to stay in the small angle region where the approximation $ x \simeq \sin (x) $ is acceptable and perform a kickback with a $Y$-rotational-gate \Cref{qc_ry_kickback}.

This matrix \acl{be} has the shape of a y-rotational-gate so that it can be used as a $\widehat{W_{Y}}$ embedded \ac{qsp} signal operator gate.
In order to exploit all the amplitude spectrum, a first \ac{qsp} with the $\mathrm{p2a}$ function \Cref{p2a_eq} linearize the cosine \Cref{p2a_explanation_eq} in the interval of interest.
Its associated phase angles are in \Cref{table_phase_angles_a}.

\begin{equation}
\begin{aligned}
    \mathrm{p2a}(x) & = \frac{2}{\pi} \cos^{-1}(x) - 1 \\
    & = -\frac{2}{\pi} \sin^{-1}(x)
    \label{p2a_eq}
\end{aligned}
\end{equation}
\begin{equation}
    \mathrm{p2a}(\cos(\theta_{a} + \Delta)) = \frac{a}{\alpha 2^{N}} + \frac{\delta}{\alpha}- 1
    \label{p2a_explanation_eq}
\end{equation}

\begin{figure}[tb]
\begin{subfigure}{0.5\textwidth}
    \begin{center}
        \resizebox{0.9\linewidth}{!}{\includegraphics{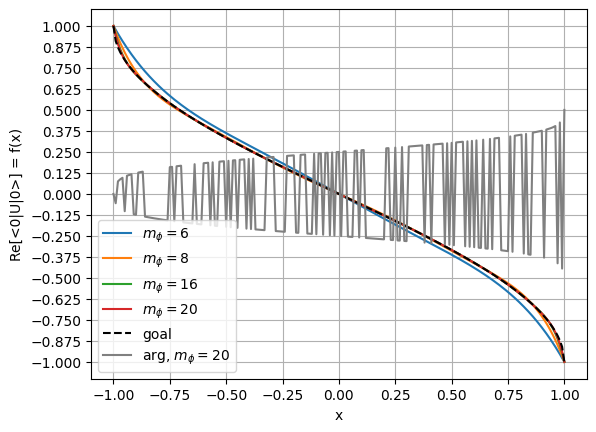}}
    \end{center}
    \caption{Approximation of the $\mathrm{p2a}$ function with respect to the state amplitude function at different degree $m_{\phi}$. \\ \\}
    \label{p2a_function}
\end{subfigure}
\begin{subfigure}{0.5\textwidth}
    \begin{center}
        \resizebox{0.9\linewidth}{!}{\includegraphics{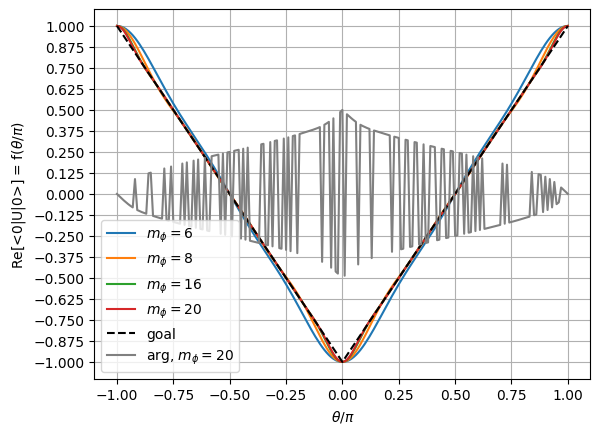}}
    \end{center}
    \caption{Approximation of the $\mathrm{p2a}$ function with respect to the state eigen-phase function at different degree $m_{\phi}$.
    The amplitude image of a phase with respect to this function will be proportional to the input phase.}
    \label{p2a_function_t}
\end{subfigure}
\end{figure}

The choice of the subspace in which the amplitudes are projected on depends on our function of interest and is controlled by two parameters, \Cref{table_function_subspace}.
When processing the quantum phase, it is essential to keep in mind that the \ac{qsp} semantic embedding have imposed value for $ x \in [ -1, 0, 1 ] $ and that working in a smaller space can leverage it.
It is a reason why working on subdomains of $x$ is interesting.
The other reason is parity constraints, which is addressed in \Cref{parity_constraints}.

\begin{table}[tb]
    \caption{Some useful subspaces to begin the phase processing thanks to embedded \ac{qsp}}
    \begin{center}
        % \resizebox{\linewidth}{!}{
            \begin{tabular}{|c|c|c|}
            \hline
            \textbf{$x$ Subspace} & \textbf{shift: $\delta$} & \textbf{scalling coef: $\alpha$} \\
            \hline
            $ [ -1; 1 ] $ & $0$ & $1$ \\
            \hline
            $ [ -1; 0 ] $ & $0$ & $2$ \\
            \hline
            $ [ 0; 1 ] $ & $1$ & $2$ \\
            \hline
            $ [ -\frac{1}{2}; \frac{1}{2} ] $ & $\frac{1}{2}$ & $2$ \\
            \hline
            $ [ \frac{1}{4}; \frac{3}{4} ] $ & $\frac{5}{2}$ & $4$ \\
            \hline
            \end{tabular}
        % }
        \label{table_function_subspace}
    \end{center}
\end{table}

\subsubsection{Amplitude Proceeding}
\begin{figure}[tb]
    \begin{center}
        \resizebox{10 cm}{!}{\includegraphics{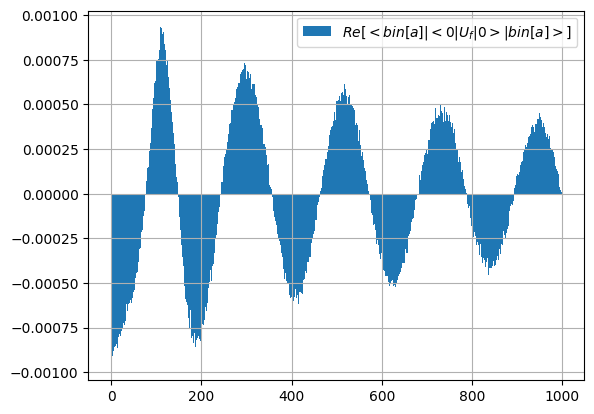}}
    \end{center}
    \caption{The real part of the quantum amplitudes resulting from the \ac{qsp} with the phase angles resulting from the merging of $\mathrm{p2a}$ and $\mathrm{f}$: \Cref{qso_qc} without $\mathrm{sgn}$ function.
    This histogramm amplitudes are obtained thanks to a Hadamard test.
    The \ac{qsp} is run on a uniform superposition of the register $\mathcal{N}$.
    The resulting amplitudes of the real part are thus proportional to the \ac{qsp} function image of $\mathcal{N}$ register associated value.
    Here: $N = 10$, $\delta = 0$ and $\alpha = 2$.
    The number of states is the number of possible binary representations $\mathrm{max}[\mathrm{bin}[a]] = 2^{N} = 1024$ and $\mathrm{max}[\mathrm{Re}[\bra{\mathrm{bin}[a]}\bra{0} \widehat{U} \ket{0} \ket{\mathrm{bin}[a]}]] = \frac{1}{2^{N}} \simeq 0.00097$.}
    \label{exemple_function_qse_p2a}
\end{figure}

This part directly depends on the \acl{qa} function of interest.
Here, the function $\mathrm{f}$ is used as an example, \Cref{exemple_function_qse_p2a}.
The unitarity of the matrix processed by \ac{qsp} imposes that the function must be normalized and evaluated on an interval between $ [ -1, 1 ] $:

\begin{equation}
    \mathrm{f} : \{ -1 ; 1 \} \longrightarrow \{ -1 ; 1 \}
    \label{funct_normalised_eq}
\end{equation}

The example's set of phase angles can be found in \Cref{table_phase_angles_a}.

This method does not allow direct quantum state initialization because the amplitudes are block-encoded.
As in \cite{mcardle_quantum_2022}, the amplitude can be amplified with \ac{eaa} to construct a state-initialization.

\subsubsection{Oracle Construction: Amplitudes Flagging}
\begin{figure}[tb]
    \begin{center}
        % \resizebox{\linewidth}{!}{
        \includegraphics{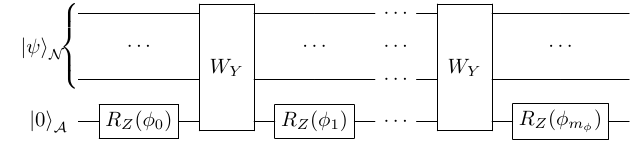}%}
    \end{center}
    \caption{The developed \acl{qso} \underline{or} $\widehat{W_{Y}}_{H_{\mathrm{f}}}$ quantum circuit.
    It uses \Cref{qc_ry_kickback} with $\underline{\phi}$ the phase angle merging following \Cref{phase_angle_merge_eq} $\mathrm{p2a}$, $\mathrm{funct}$ and $\mathrm{sgn}$ for the oracle and only $\mathrm{p2a}$ and $\mathrm{funct}$ for $\widehat{W_{Y}}_{H_{\mathrm{f}}}$.}
    \label{qso_qc}
\end{figure}

\begin{figure}[tb]
\begin{subfigure}{0.5\textwidth}
    \begin{center}
        \resizebox{0.9\linewidth}{!}{\includegraphics{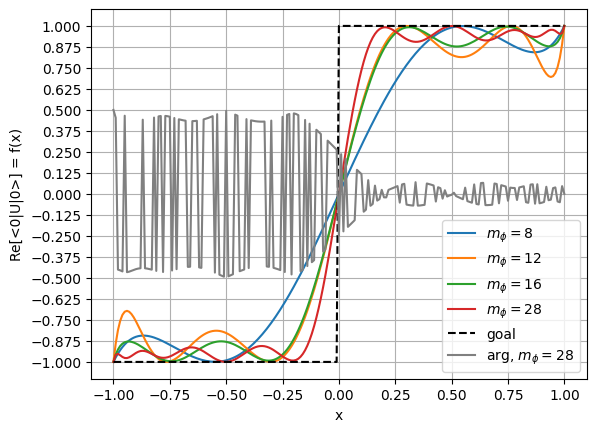}}
    \end{center}
    \caption{Approximation of the $\mathrm{sgn}$ function with respect to the state amplitude function at different degrees $m_{\phi}$. \\}
    \label{qso_function}
\end{subfigure}
\begin{subfigure}{0.5\textwidth}
    \begin{center}
        \resizebox{0.9\linewidth}{!}{\includegraphics{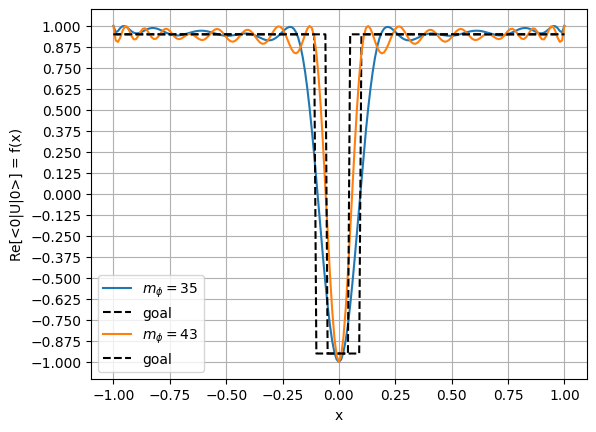}}
    \end{center}
    \caption{Approximation of the $\mathrm{step}_{1/20}$ (orange) and $\mathrm{step}_{1/10}$ (blue) function with respect to the state amplitude function.}
    \label{qso_step_function}
\end{subfigure}
\end{figure}

This section combines the tools to construct a Grover oracle that flags all the state-respecting constraints.
This is done in two steps:
\begin{itemize}
    \item Construct a quantum state that contains the function of interest thanks to $\mathrm{p2a}$ and the phase angles associated with the function of interest. The corresponding quantum circuit is illustrated in \Cref{qso_qc}.
    \item Use a filtering function that projects the value of interest to $-1$ and the other to $1$. 
    For instance:
    \begin{itemize}
        \item Sign-function \Cref{sign_function_eq} allows to seek all the function's positive (or negative) values \Cref{qso_function}.
        The phase angles of different degrees can be found in \Cref{table_phase_angles_a}.
        \item Step-function \Cref{step_function_eq} allows to seek the roots of the function values of the function \Cref{qso_step_function}.
        The phase angles of different degrees can be found in \Cref{table_phase_angles_a}.
    \end{itemize}
\end{itemize}

\begin{equation}
    \mathrm{sgn}(x) = \frac{x}{|x|}
    \label{sign_function_eq}
\end{equation}
\begin{equation}
    \mathrm{step}_{\delta_{s}}(x) = -1 \text{ if } |x| < \delta_{s} ; 1 \text{ else}
    \label{step_function_eq}
\end{equation}

\begin{figure}[tb]
    \begin{center}
        \resizebox{10 cm}{!}{\includegraphics{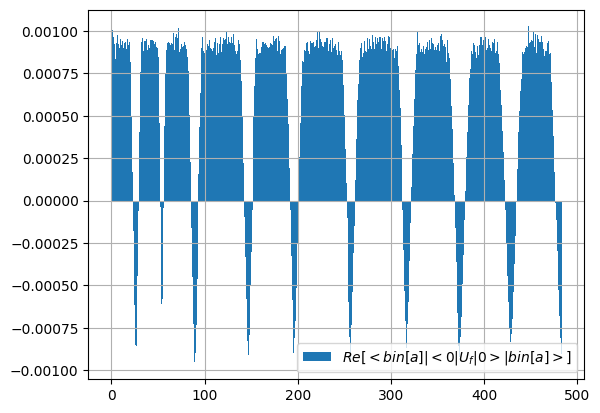}}
    \end{center}
    \caption{Example of oracle generation to mark the roots of our example function.
    This diagram was generated with the same strategy as \Cref{exemple_function_qse_p2a} with the extra embedding of the $\mathrm{step}_{1/10}$ function.}
    \label{qso_function_ex}
\end{figure}

The scaling of this technique is detailed in \Cref{table_qse_scalling} under the name of \ac{qso}.

\subsubsection{Amplitudes to Binary Register}
\begin{figure}[tb]
\begin{subfigure}{0.5\textwidth}
    \begin{center}
        \resizebox{0.9\linewidth}{!}{\includegraphics{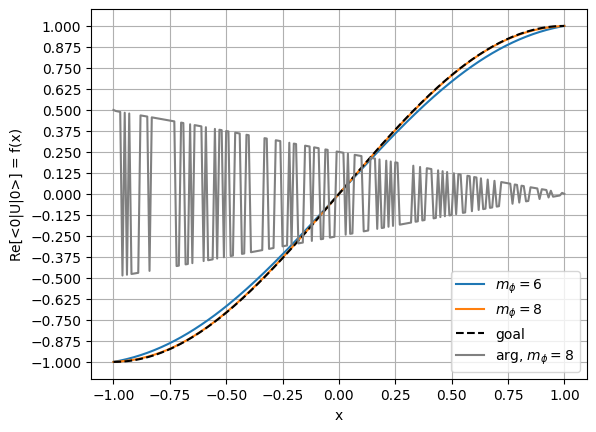}}
    \end{center}
    \caption{Approximation of the $\mathrm{a2p}$ function with respect to the state amplitude function at different degrees $m_{\phi}$.
    The phase (grey) resulting from an amplitude image will be proportional to this amplitude.}
    \label{a2p_function}
\end{subfigure}
\begin{subfigure}{0.5\textwidth}
    \begin{center}
        \resizebox{0.9\linewidth}{!}{\includegraphics{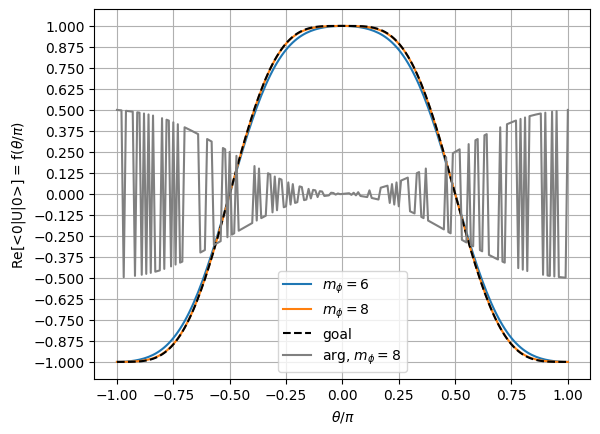}}
    \end{center}
    \caption{Approximation of the $\mathrm{a2p}$ function with respect to the state eigen-phase function at different degree $m_{\phi}$.}
    \label{a2p_function_t}
\end{subfigure}
\end{figure}

In a fashion similar to how \ac{qae} is used, reading the phase of a $\widehat{R}_{X}$ or $\widehat{R}_{Y}$ gate results in measuring an arcsine function of its amplitude \cite[Section 4: p.18]{brassard_quantum_2002}.
Since the goal is not to read the arcsine but the amplitude value, an important step is to convert the amplitude in a phase which is  proportional to this amplitude before reading the previous \ac{qsp}.
This is done with the amplitude to phase function (a well-parametrized sine) \Cref{a2p_eq}\footnote{It is interesting to note that $\mathrm{a2p}$ is the $\mathrm{p2a}$ inverse function and vice versa.}, \Cref{a2p_function} and \Cref{a2p_function_t}.

It results in reading the polynomial approximation of the function in a specific binary format.

\begin{equation}
    \mathrm{a2p}(x) = \sin(\frac{\pi}{2} x)
    \label{a2p_eq}
\end{equation}
\begin{equation}
    \pm \frac{\cos^{-1}(\mathrm{a2p}(x))}{2 \pi} = \pm \frac{1}{4} (1 - x) \text{ for } x \in [-1, 1]
    \label{a2p_explanation_eq}
\end{equation}

The phase angles associated to this function are also in the \Cref{table_phase_angles_a}.

\begin{figure}[tb]
    \begin{center}
        \resizebox{11 cm}{!}{\includegraphics{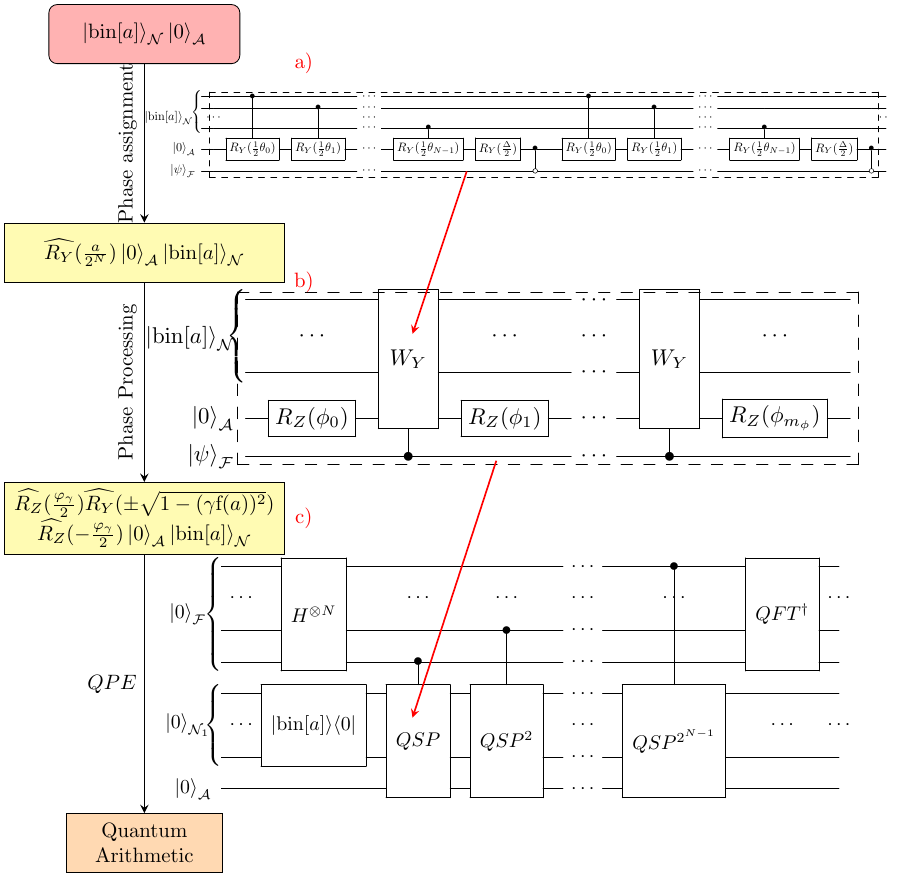}}
    \end{center}
    \caption{{\color{red} c)} Full restricted parity \ac{qse} quantum circuit.
    {\color{red} b)} Controlled-\ac{qsp} needed for the \ac{qse}.
    {\color{red} a)} $\widehat{CW_{Y}}$ controlled signal operator.
    The control qubit is the one of the $\mathcal{F}$ register.
    It exploid $ \widehat{I} = \widehat{W_{Y}} \widehat{W_{Y}}^{\dag} $ and $ \widehat{W_{Y}}^{\dag} = \widehat{Z} \widehat{W_{Y}} \widehat{Z} $.
    }
    \label{qc_final_qpe}
\end{figure}

The last step consist in reading the embedded \ac{qsp} frequecies with a \ac{qpe} algorithm \Cref{qc_final_qpe}.
A simple trick is used to control the \ac{qsp}.
An alternative solution to control the \Cref{qc_final_qpe} \ac{qsp} blocks is to switch from embedded \ac{qsp} to embedded \ac{qsvt} \cite[Section III]{rossi_semantic_2023}.
Embedded \ac{qsvt} can then simply be controlled by controlling the signal processor gate $\widehat{R_{Z}}$ and thus divide by two the cost of the circuit.

It is also important to note the numerical error due to the $\mathrm{p2a}$, $\mathrm{sgn}$ or $\mathrm{a2p}$ function polynomial approximation.
It explains the shift of the first wave of \Cref{exemple_function_qse_p2a_a2p}.
This numerical error is not evaluated in this article.
It is nevertheless possible to guess that the error is due mainly due to the arcsin function, which is hard to approximate when $ |x| > \frac{1}{2} $ \cite[Appendix D]{haner_optimizing_2018}.
It can be avoided by choosing different intervals of computing \ref{table_function_subspace}.

\begin{figure}[tb]
    \begin{center}
        \resizebox{11 cm}{!}{
        \includegraphics{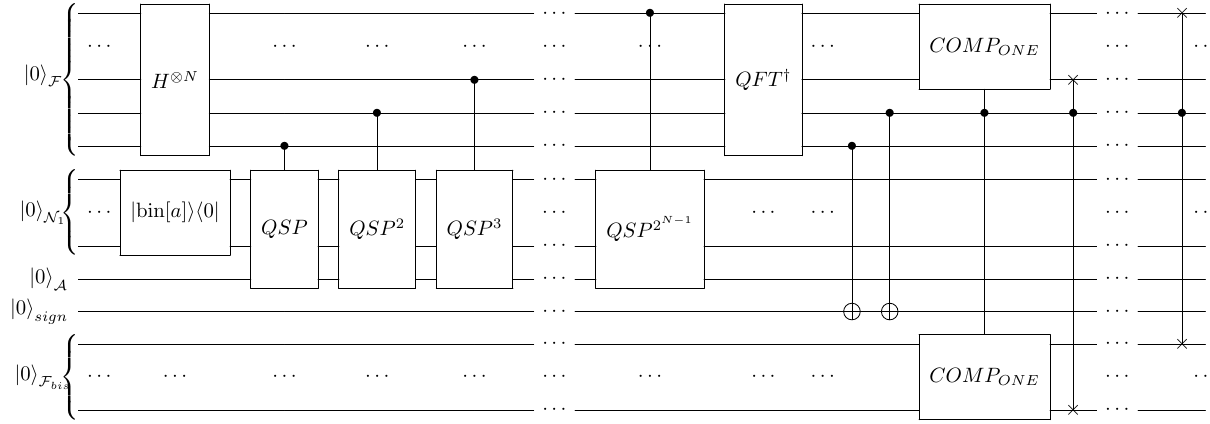}}
    \end{center}
    \caption{Full restricted parity \ac{qse} quantum circuit with the number in an easy-to-read format in the $\mathcal{F}_{bis}$ register.
    $F_{bis} = F - 1$}
    \label{qc_final_qpe_bis}
\end{figure}

\begin{figure}[tb]
    \begin{center}
        \resizebox{10 cm}{!}{
        \includegraphics{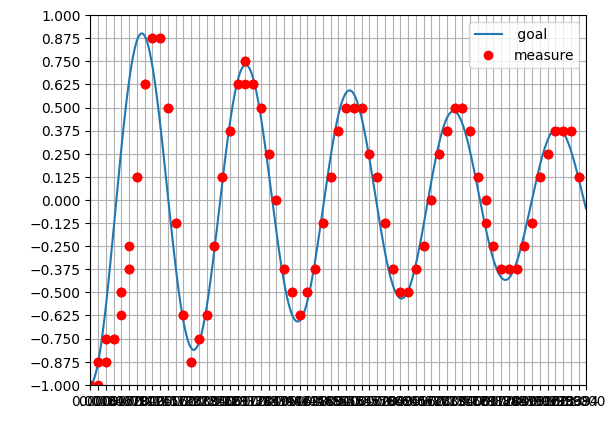}}
    \end{center}
    \caption{Binary results are read in the $\mathcal{F}$ register and translated in the usual format thanks to \Cref{qc_final_qpe_bis}.
    Here $F = 5$ and $N = 6$.}
    \label{exemple_function_qse_p2a_a2p}
\end{figure}

It is essential to highlight that the binary output will be in a specific format.
This format is determined from the value of the two most significant qubits of the $\mathcal{F}$ register.
The first indicates the sign of the phase, and the second indicates whether the value is higher or smaller than zero in the amplitude domain, respectively smaller or higher than one-fourth.
To transform the result in an easy-to-read format\footnote{In this work easy-to-read format means positive integer representation plus a bit for the sign.}, it is required to modify it depending on the area in which the phase is read:
\begin{itemize}
    \item If the phase lands in the bottom-left area of \Cref{a2p_function}: The first two qubits of the $\mathcal{F}$ register read $1 1 \mathrm{b}$ and can be ignored since one-fourth of the easy-to-read format is natively read.
    \item If the phase lands in the top-left area of \Cref{a2p_function}: The first two qubits of the $\mathcal{F}$ register read $0 1 \mathrm{b}$ and a minus sign is implicit since minus one-fourth of the easy-to-read format is natively read.
    \item If the phase lands in the top-right area of \Cref{a2p_function}: The first two qubits of the $\mathcal{F}$ register read $0 0 \mathrm{b}$, and the value read on the rest of the register is the one complement of the value in the easy-to-read format.
    \item If the phase lands in the bottom-right area of \Cref{a2p_function}: The first two qubits of the $\mathcal{F}$ register read $1  0 \mathrm{b}$, and the value read on the rest of the register is minus the one complement of the value in the easy-to-read format.
\end{itemize}
The \Cref{qc_final_qpe_bis} indicates how to translate this result in the more straightforward binary representation.
The read output is presented in the \Cref{exemple_function_qse_p2a_a2p}.

\subsubsection{Scaling and Comparison}
The results of \cite{haner_optimizing_2018}, which construct function polynomial approximations are compared to our results.
\cite{haner_optimizing_2018} use binary logic to construct the function approximations.
All the \cite{haner_optimizing_2018} registers have the same number of bits ($ N = F $) to make the implementation consistent.
In their analysis, the figures of merit used are the number of qubits and the count of Toffolly gates, which is representative of the total gate count and of its depth.
\cite[Appendix B]{haner_optimizing_2018} shows that the Toffolly gate count increases linearly with the polynomial degree $p_{\mathrm{f}}$ if the number of ancilla registers equals the polynomial degree plus one, which means $ (p_{\mathrm{f}} + 1) N $ extra qubits.
They also explain that a tradeoff between the polynomial degree and the number of ancilla registers exists for a given gate count (and so circuit depth) \cite[Table 1]{haner_optimizing_2018} and that seeking this optimal gate count is a complex problem.
It is also interesting to mention that the authors propose techniques to accelerate the computing (decreased the circuit depth) thanks to extra qubits and gates.

The scaling of the \ac{qsp} based technique is detailed in \Cref{table_qse_scalling} under the name of \ac{qse}.
The table shows that this technique gives better results regarding the qubit number (optimal plus one ancilla) to achieve any polynomial function.
It also gives interesting results regarding the number of multiqubit gate scaling with respect to the polynomial degree which is also linear in $p_{\mathrm{f}}$: $ m_{\phi} = 2 p_{\mathrm{p2a}} * 2 p_{\mathrm{a2p}} * 2 p_{\mathrm{f}} $.
The drawbacks are an overhead due to $\mathrm{a2p}$ plus $\mathrm{p2a}$ polynomial approximation and a prefactor of the linear scaling (with respect to the polynomial degree) that scales exponentially with the number of digits of the result ($F$).

\begin{table*}[tb]
    \caption{Scaling of the \ac{qso} and \ac{qse} for a fully connected quantum computer.
    The circuit depth supposes that the two-qubit gate decomposition proposed in the table is used.
    The \ac{qse} depth does not consider the \ac{qft} depth, which may vary depending on its implementation.}
    \begin{center}
        \resizebox{\linewidth}{!}{
        \begin{tabular}{|c|c|c|c|c|c|c|c|}
        \hline
        \multirow{2}{*}{\textbf{Operator}} & \multirow{2}{*}{\textbf{Qubit number}} & \multirow{2}{*}{\textbf{Circuit Depth}} & \multicolumn{5}{c|}{\textbf{Gate number}} \\
        \cline{4-8}
        & & & $\widehat{QFT}_{F}$ & $\widehat{R_{Z}}$ & $\widehat{R_{Y}}$ & $\widehat{CZ}$ & $\widehat{CR_{Y}}$ \\
        \hline
        $\widehat{W_{Y}}$ & $N + 1$ & $ N + 1 $ & $0$ & $0$ & $1$ & $0$ & $N$ \\
        \hline
        $\widehat{QSO}$ & $N + 1$ & $ m_{\phi} (N + 2) + 1 $ & $0$ & $m_{\phi} + 1$ & $m_{\phi}$ & $0$ & $m_{\phi} N$ \\
        \hline
        $\widehat{CW_{Y}}$ & $N + 2$ & $ 2 (N + 2) $ & $0$ & $0$ & $2$ & $2$ & $2 N$ \\
        \hline
        $\widehat{CQSP}$ & $N + 2$ & $ m_{\phi} (2 (N + 2) + 1) + 1 $ & $0$ & $ m_{\phi} + 1$ & $2 m_{\phi}$ & $2 m_{\phi}$ & $2 m_{\phi} N$ \\
        \hline
        $\widehat{QSE}$ & $F + N + 1$ & $ 2^{F + 1} (m_{\phi} (2 (N + 2) + 1) + 1) $ & $2$ & $2^{F+1} (m_{\phi} + 1)$ & $2^{F+2}  m_{\phi}$ & $2^{F+2} m_{\phi}$ & $2^{F+2} m_{\phi} N$ \\
        \hline
        \end{tabular}}
        \label{table_qse_scalling}
    \end{center}
\end{table*}

\subsubsection{\acl{qse} with Arbitrary Parity}
\label{parity_constraints}
To construct a function without parity constraints, a weighted sum of two \ac{qsp} (with odd and even function) results can be done (thanks to the LCU technique) in a new qubit register.
An arbitrary parity function can also be achieved by choosing an interval two times smaller when going from phase to amplitude so that only one side of the function contains all the function results.
This second option is the one used to illustrate the \ac{qse} technique in this article \Cref{exemple_function_qse}, \ref{p2a_function_t}, \ref{qso_function_ex} and \ref{exemple_function_qse_p2a_a2p}.

\section{Conclusion}
This paper presents techniques to construct a \acl{qa} circuit that is not based on binary logic.
Although this technique was initially proposed for pedagogical purposes, demonstrating the underlying principles of various quantum computing methods, it results in outcomes that differ significantly from those achieved through binary logic.
It could be efficient for specific implementations such as high-order polynomial approximation computed on its input with many digits but with a result expressed with a small number of digits due to the exponential reading overhead.
The value on the first digits will not be impacted because the computing aspect and reading of the result are separated.

This work brings a simple geometric interpretation of some quantum computing routines.
It explores the equivalence between the different queries to implement quantum information.
In particular, it shows how quantum information can be switched from phase to binary domain to be processed and gives a straightforward interpretation of the \ac{qft} adder.
It also provides a straightforward explaination why $\mathrm{a2p}$ and $\mathrm{p2a}$ functions appear natively in quantum routines such as \ac{qae} or \ac{aqe} (in \ac{hhl}) when phase to amplitude transition is needed.

\bibliographystyle{quantum}
\bibliography{biblio}

\section{Funding}
This research work was supported in part by the French PEPR integrated project Etude de la PIle Quantique — EPIQ, (ANR-22-PETQ-0007).

% \onecolumn
\appendix

\section{Appendix: Phase Angle Series}
\label{have_to_mention}
Efficient ways to search set of phase angles is an active subfield \cite{chao_finding_2020, martyn_grand_2021, dong_efficient_2021, dong_robust_2023,mizuta_recursive_2024, yamamoto_robust_2024,wang_quantum_2023, motlagh_generalized_2024}.
Even if analytical methods exist for a few functions, classical optimization is the straightforward way to seek the phase angles.
The set of phase angles proposed in the following appendix was computed using \ref{optim_funct_eq} with a parity constraint imposed on the phase angles and thanks to classical optimization.
Instead of working in the amplitude domains, it is possible to directly optimize\footnote{
    It is important to remember that this section deals with classical optimization.
    This process 'only' needs to compute $d m_{\phi}$ two-by-two matrix product for each function evaluation, which is tractable with a classical computer.
    $ d > m_{\phi} $ and the higher $d$, the better.
} a function of the phases themselves \Cref{optim_funct_arg_eq} instead of its associated amplitudes \Cref{optim_funct_eq}:

\begin{equation}
    \mathcal{L}(\underline{\phi}) = \sum_{j = 0}^{d} | \mathrm{Re}[ \bra{0} \widehat{U}_{\phi}(x_{j}) \ket{0} ] - \mathrm{f}(x_{j}) | \text{ with } x_{j} = \frac{j}{d}
    \label{optim_funct_eq}
\end{equation}
\begin{equation}
    \mathcal{L}(\underline{\phi}) = \sum_{j = 0}^{d} | \mathrm{Arg}[ \bra{0} \widehat{U}_{\phi}(x_{j}) \ket{0} ] - \mathrm{f}(x_{j}) | \text{ with } x_{j} = \frac{j}{d}
    \label{optim_funct_arg_eq}
\end{equation}

The main problem of this technique is that little is known on how to constrain the phase angles properties to seek direct phase functions that respect the embedding properties.
Other \ac{qsp} inspired structures, such as \ac{qpp} \cite{wang_quantum_2023} or \ac{gqsp} \cite{motlagh_generalized_2024}, could be used; it is nevertheless very likely that it would not allow embedding.
The function must thus be computed in one shot.

The following \Cref{table_phase_angles_a} gives the article's function antisymetric set of phase angles.

\begin{table*}[!htbp]
    \caption{Set of phase angles associated with relevant functions. The functions are from amplitude to amplitude.
    }
    \begin{center}
        \resizebox{\linewidth}{!}{
        \begin{tabular}{|c|c|p{15 cm}|}
        \hline
        \textbf{Function} & \textbf{Degree} $m_{\phi}$ & \textbf{Phase Angles} \\
        \hline
        $\mathrm{p2a}$ & $2 * 3$ & 1.220677134473429$ + \frac{\pi}{2}$, -0.9141986745515728, -1.2955052748350293, 1.2955052748350293, 0.9141986745515728, -1.220677134473429$ + \frac{\pi}{2}$ \\
        \cline{2-3}
        & $2 * 4$ & 0.8681682322909842$ + \frac{\pi}{2}$, 0.4424429404278441, 1.7226820955233317, -0.28367943409333685, 0.28367943409333685, -1.7226820955233317, -0.4424429404278441, -0.8681682322909842$ + \frac{\pi}{2}$ \\
        \cline{2-3}
        & $2 * 8$ & 1.2320580637283647$ + \frac{\pi}{2}$, 0.8252473731446527, -1.1775425298146687, 0.2540914369944401, -1.899907730573736, 0.6454484715597536, -0.4868827734973951, 1.8131545375465181, -1.8131545375465181, 0.4868827734973951, -0.6454484715597536, 1.899907730573736, -0.2540914369944401, 1.1775425298146687, -0.8252473731446527, -1.2320580637283647$ + \frac{\pi}{2}$ \\
        \hline
        & $2 * 10$ & 1.3329924058101739$ + \frac{\pi}{2}$, 1.035009436993698, 0.3490884010660929, -0.3586872071383961, -0.5059414117977128, 1.0391369102079064, -1.642759890740595, -1.1252631786120018, 0.33295057920975346, 0.6452181263562808, -0.6452181263562808, -0.33295057920975346, 1.1252631786120018, 1.642759890740595, -1.0391369102079064, 0.5059414117977128, 0.3586872071383961, -0.3490884010660929, -1.035009436993698, -1.3329924058101739$ + \frac{\pi}{2}$ \\
        \hline
        $\mathrm{a2p}$ & $2 * 3$ & 1.3276443279747514, 1.415148148001756, 1.058648479211631, -1.058648479211631, -1.415148148001756, -1.3276443279747514 \\
        \cline{2-3}
        & $2 * 4$ & 2.318411389939721, -1.328581845947473, -0.8059565062780474, -1.3599512330439159, 1.3599512330439159, 0.8059565062780474, 1.328581845947473, -2.318411389939721 \\
        \hline
        $\mathrm{sgn}$ & $2 * 4$ & -0.2640092131936705, 2.18156267897504, 2.245977049120987, -1.811175086662138, 1.811175086662138, -2.245977049120987, -2.18156267897504, 0.2640092131936705 \\
        \cline{2-3}
        & $2 * 8$ & 1.2302357850991548, 1.0336749243518562, 0.11198067732520582, -0.15680734869559498, -0.9578602870885039, 2.0612195475315174, 0.8243803424562862, -0.6554966690623732, 0.6554966690623732, -0.8243803424562862, -2.0612195475315174, 0.9578602870885039, 0.15680734869559498, -0.11198067732520582, -1.0336749243518562, -1.2302357850991548 \\
        \cline{2-3}
        & $2 * 14$ & 0.27882619783038065, 0.36281454618398395, -1.0026301277387055, 0.3541147797152542, 1.6243661292276912, 0.7417194190156123, -1.156030172384029, 1.586825848651674, -0.5126707526970495, 0.26480940875731623, -0.7942611570978513, -0.8276672490175913, 0.3724217902363619, 1.6407210513498245, -1.6407210513498245, -0.3724217902363619, 0.8276672490175913, 0.7942611570978513, -0.26480940875731623, 0.5126707526970495, -1.586825848651674, 1.156030172384029, -0.7417194190156123, -1.6243661292276912, -0.3541147797152542, 1.0026301277387055, -0.36281454618398395, -0.27882619783038065 \\
        \hline
        $\mathrm{step}_{1/20}$ & $2 * 21$ & -1.5382973389338346$ + \frac{\pi}{2}$, 2.851477247855508, -0.40984657825729826, -0.5646293285269941, -0.5972560381469758, -0.3918423962119306, 0.055463135947150695, -0.29944961000660675, -2.7921501174003267, -0.03820994975986922, -2.098844136925592, 0.5999852992281652, -0.8290246226574786, 0.018394471278165098, -0.3193690718221374, 2.8179480550390674, 2.656738804493763, -0.26716122886345256, 2.6371127484302086, -0.22243978812718435, -0.07320162044498352, 0.0, 0.07320162044498352, 0.22243978812718435, -2.6371127484302086, 0.26716122886345256, -2.656738804493763, -2.8179480550390674, 0.3193690718221374, -0.018394471278165098, 0.8290246226574786, -0.5999852992281652, 2.098844136925592, 0.03820994975986922, 2.7921501174003267, 0.29944961000660675, -0.055463135947150695, 0.3918423962119306, 0.5972560381469758, 0.5646293285269941, 0.40984657825729826, -2.851477247855508, 1.5382973389338346$ + \frac{\pi}{2}$ \\
        \hline
        $\mathrm{step}_{1/10}$ & $2 * 17$ & 0.3359251571426454$ + \frac{\pi}{2}$, -0.7053359438019324, -0.2460812016315209, 0.6290659385757417, 0.26513895221650835, 0.9253504197223081, 0.29770494357694177, 0.3113205753649513, -0.08150412617230154, -1.0429304845709697, 0.6119290859635423, 0.2688690150870127, -1.0037188566241908, -0.5852612343222781, 0.2002510054902895, -0.6262504757168341, -1.0305371577280382, 0.0, 1.0305371577280382, 0.6262504757168341, -0.2002510054902895, 0.5852612343222781, 1.0037188566241908, -0.2688690150870127, -0.6119290859635423, 1.0429304845709697, 0.08150412617230154, -0.3113205753649513, -0.29770494357694177, -0.9253504197223081, -0.26513895221650835, -0.6290659385757417, 0.2460812016315209, 0.7053359438019324, -0.3359251571426454$ + \frac{\pi}{2}$ \\
        \hline
        $\mathrm{f}$ & $2 * 22$ & -0.7919265959333694, -1.706573438268197, -0.06300361777976618, 0.14616034291649688, -0.945878016693397, -0.9831865238632601, 0.2723337690220768, -0.053752081558436623, -0.6855847965700638, -0.5029559986160452, -0.8898394054092426, 1.2477726224572974, -1.3842149465999596, -0.14604176338531175, -0.4717698081520115, 0.31556629410244413, -0.18845471639572597, -0.2160277263267179, -0.2530703519014686, -0.6397299812266477, -0.4767079419048785, -0.06737297231596824, 0.06737297231596824, 0.4767079419048785, 0.6397299812266477, 0.2530703519014686, 0.2160277263267179, 0.18845471639572597, -0.31556629410244413, 0.4717698081520115, 0.14604176338531175, 1.3842149465999596, -1.2477726224572974, 0.8898394054092426, 0.5029559986160452, 0.6855847965700638, 0.053752081558436623, -0.2723337690220768, 0.9831865238632601, 0.945878016693397, -0.14616034291649688, 0.06300361777976618, 1.706573438268197, 0.7919265959333694 \\
        \hline
        \end{tabular}}
        \label{table_phase_angles_a}
    \end{center}
\end{table*}

\newpage

\section{Annex: Link with other Quantum Routines}
As presented in the main paper, the $\mathrm{a2p}$ and $\mathrm{p2a}$ functions are already known and used in other quantum computing routines.
The following section gives more details on these routines.

\subsection{\acl{aqe}: large angle correction}
The \ac{aqe}, corrected with a $\mathrm{p2a}$ donne by \ac{qsp}, used to construct the $\widehat{W_{Y}}$ signal operator is similar to the one used in the \ac{hhl} algorithm and in state initialization \cite{mcardle_quantum_2022}.
It only differs from the \ac{aqe} presented in \Cref{section_hhl} by the techniques used to implement the function and the qubit which controls the \acl{be}.
The correction allows us to get rid of the small angle approximation ($\sin(x) = x$) needed when \ac{aqe} is used.

Convert superposition of binary numbers in \acl{be} amplitude encoding is a problem that happens in other quantum routines (such as \ac{qml} or state initialization).
It is interesting to note that it may be cheaper to use the \ac{aqe} with $\mathrm{p2a}$ function based on embedded \ac{qsp} instead of the \ac{aqe} based on logical operations for binary to amplitude conversion in other algorithms.

\subsection{Link with \acl{qae}}
In quantum computing, the Grover operator, later generalized into the \ac{qae} operator, allows to construct from arbitrary unitary $\widehat{U}$, a gate that behaves like qubitized $X/Y$-rotationnal: $\widehat{W_{X/Y}}$ gates:
\begin{equation}
    \widehat{R_{X/Y}}(\theta, \varphi) =
    \begin{bmatrix}
        \alpha & - \beta^{*} \\
        \beta & \alpha 
    \end{bmatrix}
    \begin{matrix}
        \ket{0} \\
        \ket{0_{\perp}}
    \end{matrix} \text{ with }
    \begin{Bmatrix}
        \alpha = \cos(\theta) \\
        \beta = \sin(\theta) e^{i \varphi}
    \end{Bmatrix}
    \label{eq_ry}
\end{equation}
The initial Grover operator transform, an arbitrary $SU(2)$, the initial state expressed (or projected) in the good state basis:
\begin{equation}
\begin{aligned}
    & \quad \begin{matrix}
        \qquad \ket{\mathit{good}} & \qquad \ket{\mathit{good}_{\perp}}
    \end{matrix} \\
    \widehat{U}
    & = \begin{bmatrix}
        \braket{\mathit{good}|\psi_{i}} & \braket{\mathit{good}_{\perp}|\psi_{i}} \\
        \braket{\mathit{good}|\psi_{i \perp}} & \braket{\mathit{good}_{\perp}|\psi_{i \perp}}
    \end{bmatrix}
    \begin{matrix}
        \ket{\psi_{i}} \\
        \ket{\psi_{i \perp}}
    \end{matrix} \\
    & = \begin{bmatrix}
        \cos(\gamma) e^{i \varphi_{c}} & \sin(\gamma) e^{i \varphi_{s}} \\
        - \sin(\gamma) e^{- i \varphi_{s}} & \cos(\gamma) e^{- i \varphi_{c}} 
    \end{bmatrix}
    \begin{matrix}
        \ket{\psi_{i}} \\
        \ket{\psi_{i \perp}}
    \end{matrix} \\
    & = \widehat{U}_{\mathit{good}} \widehat{U}_{\psi_{i}}^{\dag}
\end{aligned}
\end{equation}
into a qubitized $\widehat{W_{X/Y}}$ signal operator:
\begin{equation}
\begin{aligned}
    \widehat{G} & = \widehat{S}_{\mathit{good}} \widehat{S}_{\psi_{i}} \\
    & = \begin{bmatrix}
        -1 & 0 \\
        0 & 1
    \end{bmatrix}
    \begin{bmatrix}
        1 - 2 | \cos(\gamma) |^{2} & 2 \sin(\gamma) \cos(\gamma) e^{i \varphi} \\
        2 \sin(\gamma) \cos(\gamma) e^{- i \varphi} & 1 - 2 | \sin(\gamma) |^{2}
    \end{bmatrix}
    \\
    & = 
    \begin{bmatrix}
        \cos(2 \gamma) & - \sin(2 \gamma) e^{i \varphi} \\
        \sin(2 \gamma) e^{- i \varphi} & \cos(2 \gamma)
    \end{bmatrix}
    \begin{matrix}
        \ket{\mathit{good}} \\
        \ket{\mathit{good}_{\perp}}
    \end{matrix} \\
    & = e^{i \frac{\varphi}{2} ( \ket{\mathit{good}} \bra{\mathit{good}} - \ket{\mathit{good_{\perp}}} \bra{\mathit{good_{\perp}}} ) } e^{i 2 \gamma (i \ket{\mathit{good}} \bra{\mathit{good_{\perp}}} + \mathit{h.c.} ) } \\
    & \qquad \qquad \qquad \qquad e^{- i \frac{\varphi}{2} ( \ket{\mathit{good}} \bra{\mathit{good}} - \ket{\mathit{good_{\perp}}} \bra{\mathit{good_{\perp}}} ) }
\end{aligned}
\end{equation}
with $ \varphi = \varphi_{c} - \varphi_{s} $ and
\begin{equation}
\begin{aligned}
    \widehat{S}_{\psi} & = \widehat{U}_{\psi} \widehat{S}_{0} \widehat{U}_{\psi}^{\dag} \\
    & = \widehat{I} - 2\ket{\psi} \bra{\psi} \\
    \widehat{U}_{\psi} & = \ket{\psi} \bra{0} + \ket{\psi_{\perp}} \bra{0_{\perp}}
\end{aligned}
\end{equation}

For such a gate, the phases and amplitudes are always linked thanks to the $\mathrm{a2p}$ and $\mathrm{p2a}$ functions.
\ac{qae} algorithm is used to compute the phase $ 2 \gamma $ of this gate thanks to \ac{qpe} from which are extracted the amplitude with the $\mathrm{p2a}$ functions\cite[Section 4: p.18]{brassard_quantum_2002}\footnote{In the original article, the authors seek the square of the amplitude which equal the probability $ | \braket{\mathit{good}|\psi_{i}} |^{2} $ to measure $\ket{\mathit{good}}$ after one application, so they use $\mathrm{p2a}^{2}$ function to convert their \ac{qpe} results.}. % ! pas and Counting  car le nombre de sollution correspond directement à la phase.
Indeed, the \ac{qae} evaluated gate is a specific Grover operator for which instead of knowing $\ket{\mathit{good}}$, $ \widehat{U} $ is known\footnote{
    Interestingly, this techniques can be extended to arbitrary inner product.
    It is done by taking advantage of: $ \widehat{U}_{\psi} \widehat{U}_{\phi}^{\dag} = \ket{\psi} \bra{\phi} + \ket{\psi_{\perp}} \bra{\phi_{\perp}} $.
    It leads to:
    \begin{equation*}
        \widehat{U} = \widehat{U_{\psi}} \widehat{U_{\phi}}^{\dag} \widehat{V} =
        \begin{bmatrix}
            \braket{\phi|\widehat{V}|\psi} & \braket{\phi_{\perp}|\widehat{V}|\psi} \\
            \braket{\phi|\widehat{V}|\psi_{\perp}} & \braket{\phi_{\perp}|\widehat{V}|\psi_{\perp}}
        \end{bmatrix}
        \begin{matrix}
            \ket{\psi} \\
            \ket{\psi_{\perp}}
        \end{matrix}
    \end{equation*}
    and
    \begin{equation*}
        \widehat{A} \widehat{S}_{0} \widehat{A}^{\dag} \widehat{S}_{\psi} = \widehat{U} \widehat{S}_{\psi} \widehat{U}^{\dag} \widehat{S}_{\psi} = \widehat{V} \widehat{S}_{\psi} \widehat{V}^{\dag} \widehat{S}_{\phi}
    \end{equation*}
}:
\begin{equation}
\begin{aligned}
    \ket{\mathit{good}} & = \widehat{U} \ket{\psi_{i}} \Leftrightarrow \widehat{U}_{\mathit{good}} = \widehat{U} \widehat{U}_{\psi_{i}} = \widehat{A} \\
    \Rightarrow & \braket{\mathit{good}|\psi_{i}}  = \bra{\psi_{i}} \widehat{U} \ket{\psi_{i}} 
\end{aligned}
\end{equation}
If $\widehat{U}$ has already a $\widehat{W}_{X/Y}$ form, it imply:
\begin{equation}
\begin{aligned}
    \widehat{A} \widehat{S}_{0} \widehat{A}^{\dag} \widehat{S}_{\psi_{i}}
    & = \widehat{U} \widehat{S}_{\psi_{i}} \widehat{U}^{\dag} \widehat{S}_{\psi_{i}} \\
    & = \widehat{U}^{2}
\end{aligned}
\end{equation}

Applying the same oracle transformation with $\widehat{U}$ the result of a \ac{qsp} in the $\widehat{W_{X/Y}}$ convention leads naturally to a phase anti-symmetrization similarly to the one explained in \cref{qsp_basics}:
\begin{equation}
\begin{aligned}
    & \quad \widehat{S}_{0_{\mathcal{A}}} \widehat{U}^{\dag} \widehat{S}_{0_\mathcal{A}} \\
    & = \widehat{S}_{0_\mathcal{A}} \widehat{R_{Z}}(- \phi_{0}) \prod_{j = 1}^{m_{\phi}} [ \widehat{W_{X/Y}}^{\dag} \widehat{R_{Z}}(- \phi_{j}) ] \widehat{S}_{0_\mathcal{A}} \\
    & = \widehat{S}_{0_\mathcal{A}} \widehat{R_{Z}}(- \phi_{0}) \widehat{S}_{0_\mathcal{A}} \prod_{j = 1}^{m_{\phi}} [ \widehat{S}_{0_\mathcal{A}} \widehat{W_{X/Y}}^{\dag} \widehat{S}_{0_\mathcal{A}} \widehat{S}_{0_\mathcal{A}} \widehat{R_{Z}}(- \phi_{j}) \widehat{S}_{0_\mathcal{A}} ] \\
    & = \widehat{R_{Z}}(- \phi_{0}) \prod_{j = 1}^{m_{\phi}} [ \widehat{W_{X/Y}} \widehat{R_{Z}}(- \phi_{j}) ] 
\end{aligned}
\end{equation}
and so:
\begin{equation}
\begin{aligned}
    & \quad \widehat{U} \widehat{S}_{0_\mathcal{A}} \widehat{U}^{\dag} \widehat{S}_{0_\mathcal{A}} \\
    & = \prod_{j = 1}^{m_{\phi}} [ \widehat{R_{Z}}(\phi_{j}) \widehat{W_{X/Y}} ]
    \prod_{j = 1}^{m_{\phi}} [ \widehat{W_{X/Y}} \widehat{R_{Z}}(- \phi_{j}) ] 
\end{aligned}
\end{equation}
which is an anti-symetric \ac{qsp} in the $\widehat{W_{X/Y}}$ convention.

\section{Annex Phase Kick-back: \acl{qa} based \acl{hs}}
The \Cref{hs_based_qa} shows how to construct \acl{qa} from the \acl{hs} of the function.
This argument can also be reversed: \acl{qa} is often the most convenient strategy to construct the \acl{hs} associated with a function.
It leads back to the first quantum algorithm proposals which were mainly based on the notion of phase kickback \cite{cleve_quantum_1998} (and \cite{ossorio-castillo_generalisation_2023} for extra algorithm description with this formalism) which general form is detailled in \cite[Section 7]{cleve_quantum_1998}.
Three phase kickbacks, which behaves like a phase gates, combined with basis change on the control register, also allows the construction of the rotation gate that appears in the Hamiltonian simulation of Hermitian matrices \cite[Section 2.1]{cody_jones_faster_2012}.
Phase kickback is easily explained by developing the \acl{qa}\footnote{
    Note that the operator must leave all the ancilla registers unchanged for a kickback to appear between the superposed state of the control register.
    The artifact or garbage that can affect the ancilla register can be removed using the following composed \acl{qa} gate: 
    \begin{equation*}
    \begin{aligned}
        & \widehat{U}_{\mathrm{f}} & \ket{\mathrm{bin}[x]} \ket{\mathrm{bin}[b]} \ket{0} \ket{0} \\
        = & \widehat{U}_{\mathrm{cf}}^{\dag} \widehat{ADDm} \widehat{U}_{\mathrm{cf}} & \ket{\mathrm{bin}[x]} \ket{\mathrm{bin}[b]} \ket{0} \ket{0} \\
        = & \widehat{U}_{\mathrm{cf}}^{\dag} \widehat{ADDm} & \ket{\mathrm{bin}[x]} \ket{\mathrm{bin}[b]} \ket{\mathrm{bin}[\mathrm{f}(x)]} \ket{\beta} \\
        = & \widehat{U}_{\mathrm{cf}}^{\dag} & \ket{\mathrm{bin}[x]} \ket{\mathrm{bin}[(\mathrm{f}(x) + b)\mathrm{mod}[2^{F}]]} \ket{\mathrm{bin}[\mathrm{f}(x)]} \ket{\beta} \\
        = & & \ket{\mathrm{bin}[x]} \ket{\mathrm{bin}[(\mathrm{f}(x) + b)\mathrm{mod}[2^{F}]]} \ket{0} \ket{0}
    \end{aligned}
    \end{equation*}
    with $\widehat{U}_{\mathrm{cf}}$ the circuit directly coming from the classical gates translation into Toffoli gates.
    This tricks comes from classical reversible computing \cite[Table.I]{bennett_logical_1973}.
} as a sum of controlled unitary whose phases are proportional to the function images of the control key:

\begin{equation}
\begin{aligned}
    \widehat{U}_{\mathrm{f}} & = \sum_{x = 0}^{2^{F}} \ket{\mathrm{bin}[x]} \bra{\mathrm{bin}[x]} \otimes \ket{\mathrm{bin}[(\mathrm{f}(x) + b)\mathrm{mod}[2^{F}]]} \bra{\mathrm{bin}[b]} \\
    & = \widehat{C^{F}}\{ \ket{\mathrm{bin}[x]} \} \otimes \widehat{ADDm}_{\mathrm{f}(x)}
\end{aligned}
\end{equation}

This unitary, controlled modular adder is a powerful tool to transform a binary value into the phase of a phase gate.
In order to control the kickback factor $l$, the adder must be applied to the vector of its eigenbasis:

\begin{equation}
\begin{aligned}
    \{ \ket{\lambda_{l}} \} = & \left\{ \sum_{x = 0}^{2^{qb_{nb}} - 1} e^{- i \frac{2 \pi}{2^{qb_{nb}}} l x} \ket{\mathrm{bin}[x]} \right\} \\
    = & \left\{ \widehat{QFT} \ket{\mathrm{bin}[l]} \right\} \\
    & \text{with } l \in \{ 0, 1, 2, ..., 2^{qb_{nb}} \}
\end{aligned}
\end{equation}

so that:

\begin{equation}
    \widehat{ADDm}_{\mathrm{f}(x)} \ket{\lambda_{l}} = e^{- i 2 \pi l \frac{\mathrm{f}(x)}{2^{qb_{nb}}}} \ket{\lambda_{l}}
\end{equation}

On the computational basis, the eigenvalues of a projection of the gate eigenstate are read.
The read eigenphase is thus proportional to the sought phase, and more accurately, the smaller eigenphase ($l = 1$) is equal to the function image.
A more straightforward way to understand it is to estimate the phase of this modular adder in a binary fashion:

\begin{equation}
\begin{aligned}
    & \widehat{I} = \ket{\mathrm{bin}[(\mathrm{f}(x) + b)\mathrm{mod}[2^{F}]]} \bra{\mathrm{bin}[b]}^{n} \\
    \Leftrightarrow & (n \mathrm{f}(x) + b)\mathrm{mod}[2^{F}] = b \\
    \Leftrightarrow & \frac{1}{n} = l \frac{\mathrm{f}(x)}{2^{F}}
\end{aligned}
\end{equation}

These properties are used by Shor's algorithm in which, since multiple of the eigenphase is seeked, a superposition of all the eigenstate except ($l = 0$) can be used:

\begin{equation}
\begin{aligned}
    & \sum_{l = 0}^{2^{F} - 1} \widehat{U}_{\mathrm{f}} \ket{\lambda_{l}} = \ket{\mathrm{bin}[\alpha]} \\
    & \text{with } \mathrm{f}(\alpha) = 0
\end{aligned}
\end{equation}

In Shor's algorithm: $ \mathrm{f}(x) = e^{x} \Rightarrow \alpha = 1 $

\section{Annex: Embeded \acl{qsp} and \acl{qpe}}

\subsection{Essential details about \acl{qsp} Semantic Embedding}
\label{qsp_basics}
The \ac{qsp} algorithm allows the processing of the quantum state amplitude real part with a polynomial function \cite{martyn_grand_2021}.
It is done by consecutive repetition of a signal operator $\widehat{W_{\sigma}}$ that encodes the problem and a signal processor $\widehat{S_{\sigma}}$ with a set of angles depending on the function of interest \Cref{qsp_eq}.
The polynomial degree is the number of repetitions $m_{\phi}$.
The parity of the polynomial is the parity of the number of repetitions.
Usually, a \ac{lcu} is done with the odd and even polynomial approximations to achieve an arbitrary polynomial.

\begin{equation}
    \begin{aligned}
        \widehat{U_{\phi}} & = \prod_{j = 1}^{m_{\phi}} [ \widehat{S_{\sigma_{2}}}(\phi_{j}) \widehat{W_{\sigma_{1}}} ] \widehat{S_{\sigma_{2}}}(\phi_{0}) \\
        & = 
        \widehat{U_{a}}
        \begin{bmatrix}
            P(a) & i Q(a) \sqrt{1 - a^{2}} \\
            i Q^{*}(a) \sqrt{1 - a^{2}} & P^{*}(a)
        \end{bmatrix}
        \widehat{U_{b}}
    \end{aligned}
    \label{qsp_eq}
\end{equation}

with: $ \widehat{U_{a}} \widehat{\sigma_{1}} \widehat{U_{b}} = \widehat{X} \text{ and } \widehat{U_{a}} \sigma_{2} \widehat{U_{b}} = \widehat{Z} $.

\begin{figure}[tb]
    \begin{center}
        % \resizebox{\linewidth}{!}{
        \includegraphics{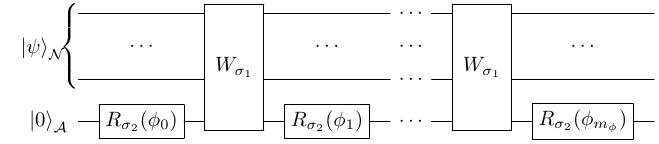}%}
    \end{center}
    \caption{Basic \ac{qsp} quantum circuit.}
    \label{qsp_qc}
\end{figure}

$P$ and $Q$ are fixed parity complex polynomials whose order are equal to $m_{\phi}$ and $m_{\phi} - 1$ respectively.
The corresponding circuit is illustrated by \Cref{qsp_qc}.

All the angles in the paper come from classical optimization \Cref{have_to_mention}.

The primary condition to process amplitudes thanks to \ac{qsp} is to encode this amplitude thanks to a signal operator qubitized with a Pauli rotation shape \Cref{qubitized_eq} and to process it with a signal processor which is perpendicular in the Bloch Sphere.
It leads to the \ac{qsp} conventions of \Cref{table_qsp_conv}.
It is relevant to remember that other conventions based on different matrices (\ac{qsvt} \cite{gilyen_quantum_2019-1}) or to different sets of phase angles (\ac{gqsp} \cite{motlagh_generalized_2024} and \ac{qpp} \cite{wang_quantum_2023}) exist.

\begin{equation}
    \begin{aligned}
        \widehat{R_{X}}(-2\cos^{-1}(a)) = 
        \begin{bmatrix}
            a & i \sqrt{I - a^{2}} \\
            i \sqrt{I - a^{2}} & a
        \end{bmatrix} &
        \begin{matrix} 
            \ket{0}_{\mathcal{A}} \\
            \ket{1}_{\mathcal{A}} \\
        \end{matrix} \\
        \widehat{W_{X}} = 
        \begin{bmatrix}
            \alpha a & i \sqrt{1 - \alpha^{2} a^{2}} \\
            i \sqrt{1 - \alpha^{2} a^{2}} & \alpha a
        \end{bmatrix} &
        \begin{matrix} 
            \ket{0}_{\mathcal{A}} \ket{\mathrm{bin}[a]}_{\mathcal{N}} \\
            \ket{1}_{\mathcal{A}} \ket{\mathrm{bin}[a]}_{\mathcal{N}} \\
        \end{matrix}
    \end{aligned}
    \label{qubitized_eq}
\end{equation}

\begin{table}[tb]
    \caption{The main \ac{qsp} conventions, the Wx-y convention operator is detailled by \Cref{rxy_gate_eq}}
    \begin{center}
        % \resizebox{\linewidth}{!}{
            \begin{tabular}{|c|c|c|c|c|}
            \hline
            \multirow{2}{*}{\textbf{Convention}} & \textbf{Signal Operator} & \textbf{Signal processor} & \multicolumn{2}{c|}{\textbf{Bases change}} \\
            & $ \widehat{W_{\sigma_{1}}} $ & $ \widehat{S_{\sigma_{2}}} $ & $ \widehat{U_{a}} $ & $ \widehat{U_{b}} $ \\
            \hline
            Wx & $ \widehat{W_{X}} $ & $ \widehat{S_{Z}} $ & \multicolumn{2}{c|}{$ \widehat{I} $} \\
            \hline
            Wy & $ \widehat{W_{Y}} $ & $ \widehat{S} \widehat{S_{Z}} \widehat{S}^{\dag} = \widehat{S_{Z}} $ & $ \widehat{S} $ & $ \widehat{S}^{\dag} $ \\
            \hline
            Wz & $ \widehat{W_{Z}} $ & $ \widehat{S_{X}} $ & \multicolumn{2}{c|}{$ \widehat{H} $} \\
            \hline
            Reflection & $ \widehat{W_{Y}} \widehat{Z} $ & $ \widehat{S}^{\dag} \widehat{S_{Z}} \widehat{S}^{\dag} = \widehat{Z} \widehat{S_{Z}} $ & \multicolumn{2}{c|}{$ \widehat{S}^{\dag} $} \\
            \hline
            Wx-y & $ \widehat{W_{XY}}_{\varphi} $ & $ \widehat{R_{Z}}(\frac{\varphi}{2}) \widehat{S_{Z}} \widehat{R_{Z}}(- \frac{\varphi}{2}) $ & $ \widehat{R_{Z}}(\frac{\varphi}{2}) $ & $ \widehat{R_{Z}}(- \frac{\varphi}{2}) $ \\
            \hline
            \end{tabular}
        % }
        \label{table_qsp_conv}
    \end{center}
\end{table}

\begin{equation}
    \begin{aligned}
        \widehat{W_{XY}}_{\varphi} & = \widehat{R_{Z}}(\varphi) \widehat{W_{X}} \widehat{R_{Z}}(- \varphi) \\
        & = 
        \begin{bmatrix}
            A & i e^{i \varphi} \sqrt{I - A^{2}} \\
            i e^{-i \varphi} \sqrt{I - A^{2}} & A
        \end{bmatrix}
    \end{aligned}
    \label{rxy_gate_eq}
\end{equation}

To process a signal operator with many layers of \ac{qsp}, the \ac{qsp} protocol final matrix must keep the shape of a signal operator.
A \ac{qsp} protocol that gives a new signal operator with a known signal processor allows to perform functions of function.

A proposal to construct a semantic embedding of \ac{qsp} by \cite{rossi_semantic_2023} is to use \ac{qsp} algorithm with antisymmetric phases angles \cite[Section 2, Definition II.3]{rossi_semantic_2023} and starting with a signal operator in the $X$-$Y$ plan of the Bloch-sphere called twisted oracle \cite[Section 2, Definition II.2]{rossi_semantic_2023}.
This antisymmetry imposes to process two times more repetitions of the signal operator to achieve the same polynomial degree.
It is enough to allows multilayer \ac{qsp} because:
\begin{itemize}
\item $\widehat{W_{XY}}_{\varphi}$ signal operators \Cref{rxy_gate_eq} are a subclass of the operator resulting from a \ac{qsp} \Cref{qsp_eq}: $ \widehat{W_{XY}}_{\varphi} \in \widehat{U_{\phi}} $ with $ P = a $ and $ Q = e^{i \varphi} $.
\item All $\widehat{W_{XY}}_{\varphi}$ signal operators can be processed with the same series of phase angles.
\subitem It can be explained because the extra rotations around the $Z$-axis $\varphi$ do not cancel each other except for the first and last rotation \Cref{qsp_eq}.
The first and last rotations do not change the amplitudes of the real part \Cref{rxy_gate_eq}.
\item The matrix resulting from the antisymmetric \ac{qsp} of a $\widehat{W_{XY}}_{\varphi}$ signal operator keeps the same shape.
As shows by \Cref{rxy_gate_eq} and \Cref{rot_rot_eq} for one iteration, it is proved by induction starting from the middle of the \ac{qsp} \Cref{qsp_induction_eq}.
\subitem Note that this proof also works for even antisymetric \ac{qsp} protocol, it begins with a different signal operator: $ \widehat{U_{\phi}}_{0} \longleftarrow \widehat{U_{\phi}}^{2} $ instead of $ \widehat{U_{\phi}}_{0} \longleftarrow \widehat{U_{\phi}} $.
\subitem It is also important to note that $P_{n}$ remains a real polynom \cite[Section 2, Definition II.3]{rossi_semantic_2023}.
\end{itemize}

\begin{equation}
    \begin{aligned}
        \widehat{U_{\phi}}_{\varphi} & = \widehat{R_{Z}}(\varphi) \widehat{U_{\phi}} \widehat{R_{Z}}(- \varphi) \\
        & = 
        \begin{bmatrix}
            P(a) & i Q(a) e^{i \varphi} \sqrt{1 - a^{2}} \\
            i Q^{*}(a) e^{- i \varphi} \sqrt{1 - a^{2}} & P(a)
        \end{bmatrix}
    \end{aligned}
    \label{rotqsp_gate_eq}
\end{equation}

\begin{equation}
    \begin{aligned}
        \widehat{U_{\phi}}_{2} & = \widehat{U_{\phi}} \widehat{U_{\phi}}_{\varphi} \widehat{U_{\phi}} \\
        & = 
        \begin{bmatrix}
            P_{2}(a) & i Q_{2}(a) \sqrt{1 - a^{2}} \\
            i Q_{2}^{*}(a) \sqrt{1 - a^{2}} & P_{2}(a)
        \end{bmatrix} \\
        & =
        \begin{bmatrix}
            P_{2}(a) & e^{i \varphi_{2}} \sqrt{1 - P_{2}(a)^{2}} \\
            e^{- i \varphi_{2}} \sqrt{1 - P_{2}(a)^{2}} & P_{2}(a)
        \end{bmatrix} &
    \end{aligned}
    \label{rot_rot_eq}
\end{equation}

with:
\begin{equation*}
    \begin{aligned}
        1 & = P^{2} + |Q|^{2} (1 - a^{2}) \\
        P_{2} & = P ( P^{2} (2 + \cos(\varphi)) - (1 + \cos(\varphi)) ) \\
        Q_{2} & = Q ( 2 P^{2} (1 + \cos{\varphi}) - e^{- i \varphi} ) \\
        e^{i \varphi_{2}} & = \frac{i Q_{2}(a) \sqrt{1 - a^{2}}}{\sqrt{1 - P_{2}(a)^{2}}}
    \end{aligned}
\end{equation*}

\begin{equation}
    \begin{aligned}
        \widehat{QSP} & = \dots \widehat{U_{\phi}} \widehat{R_{Z}}_{2} \widehat{U_{\phi}} \widehat{R_{Z}}_{1} \widehat{U_{\phi}}_{0} \widehat{R_{Z}}_{1}^{\dag} \widehat{U_{\phi}} \widehat{R_{Z}}_{2}^{\dag} \widehat{U_{\phi}} \dots \\
        & = \dots \widehat{U_{\phi}} \widehat{R_{Z}}_{2} \widehat{U_{\phi}} \widehat{U_{\phi}}_{\varphi} \widehat{U_{\phi}} \widehat{R_{Z}}_{2}^{\dag} \widehat{U_{\phi}} \dots \\
        & = \dots \widehat{U_{\phi}} \widehat{R_{Z}}_{2} \widehat{U_{\phi}}_{2} \widehat{R_{Z}}_{2}^{\dag} \widehat{U_{\phi}} \dots
    \end{aligned}
    \label{qsp_induction_eq}
\end{equation}

Instead of calling \ac{qsp} inside \ac{qsp} to produce the embedded \ac{qsp}, it is equivalently possible to merge the sets of phase angles.
It is important to recall that the new set of phase angles is not simply the previous set of phases end-to-end but follows the shape of \Cref{phase_angle_merge_eq} called phase nesting in [Table 1]\cite{rossi_semantic_2023}.
Using: $$ \underline{\phi}_{\gamma}[i, j] = ( \phi_{\gamma i}, \phi_{\gamma (i + 1)}, \dots, \phi_{\gamma (j - 1)}, \phi_{\gamma j} $$ with $ j > i $ and $ i, j \in [0, m_{\phi_{\gamma}}] ) $.

\begin{equation}
    \begin{aligned}
        \mathrm{merge}[\underline{\phi_{1}}, \underline{\phi_{2}}] = ( \phi_{1 0} + \phi_{2 0}, \underline{\phi}_{1}[1, m_{\phi_{1}} - 1], \phi_{2, 1}, \underline{\phi}_{1}[1, m_{\phi_{1}} - 1], \\
        \phi_{2 2}, \dots, \phi_{2 m_{\phi_{2} - 1}}, \underline{\phi}_{1}[1, m_{\phi_{1}} - 1], \phi_{1 m_{\phi_{1}}} + \phi_{2 m_{\phi_{2}}} )
    \end{aligned}
    \label{phase_angle_merge_eq}
\end{equation}

From this section, when we refer to \ac{qsp}, it is implicit that the set of phase angles is antisymmetric.

A last but significant remark about antisymmetric \ac{qsp} is that antisymmetry imposes extra constraints on the functions:
\begin{itemize}
\item $ P(\pm 1) = 1 $ odd antysymetric \ac{qsp}.
\item $ P(0) = \pm 1 $ odd antysymetric \ac{qsp}.
\item $ P(\pm 1) = \mp1 $ even antysymetric \ac{qsp}.
\item $ P(0) = 0 $ even antysymetric \ac{qsp}.
\end{itemize}

\subsection{Important details about \acl{qpe} of a \acl{qsp} Semantic Embeding Result}
To understand why, for the embedded \ac{qsp} result, the \ac{qpe} algorithm can be used to read the state amplitudes, let us remember that the phases read by the \ac{qpe} $ \Phi = \frac{1}{n} $ are defined as each phase that respect \Cref{qpe_result_eq}.
Only two phases can be measured in the \ac{qse} case: $\Phi$ and $-\Phi$.
The second phase associated with the other rotational direction is also measured; this phase is $-\frac{1}{n}$, which is measured as the first phase one's complement.

\begin{equation}
    \widehat{U_{b}}
    \begin{bmatrix}
        e^{i \Phi_{1}} & 0 \\
        0 & e^{i \Phi_{2}}
    \end{bmatrix}
    \widehat{U_{b}}^{\dag}
    = 
    \widehat{U_{\phi}}^{n} \text{ with }
    n_{j} \text{ so that } 1 = e^{i \Phi_{j}}
    \label{qpe_result_eq}
\end{equation}

Let us also remember that antisymetric \ac{qsp} result have a real processed signal and so a basis change: $ \widehat{U_{b}} = \widehat{R_{Z}}(\varphi) \widehat{H} $.
Its phase is linked to its amplitude by the following relation \Cref{qpe_qsp_eq}.

\begin{equation}
\begin{aligned}
    \widehat{I}
    & = 
    \begin{bmatrix}
        P & \sqrt{1 - P^{2}} \\
        \sqrt{1 - P^{2}} & P
    \end{bmatrix}^{n} \\
    & = 
    \begin{bmatrix}
        T_{n}(P) & \sqrt{1 - T_{n}(P)^{2}} \\
        \sqrt{1 - T_{n}(P)^{2}} & T_{n}(P)
    \end{bmatrix} \\
    & = 
    \begin{bmatrix}
        \cos(n \cos^{-1}(P)) & \sqrt{1 - T_{n}(P)^{2}} \\
        \sqrt{1 - T_{n}(P)^{2}} & \cos(n \cos^{-1}(P))
    \end{bmatrix} \\
    & \Leftrightarrow 2 \pi = \pm n \cos^{-1}(P) \\
    & \Leftrightarrow \frac{1}{n} = \pm \frac{\cos^{-1}(P)}{2 \pi}
\end{aligned}
\label{qpe_qsp_eq}
\end{equation}

\end{document}